\documentclass[jonas,preprint]{imsart}

\usepackage{natbib}
\usepackage{subfig}
\usepackage{booktabs}

\usepackage[latin1]{inputenc} 
\usepackage[british]{babel}

\usepackage{latexsym}
\usepackage[dvips]{graphics,graphicx,psfrag}

\usepackage{amssymb,amsmath,amsfonts,amsthm}
\usepackage{graphics}

\usepackage{verbatim}
\usepackage{pstricks,pst-node,psfrag}
\usepackage{epsfig}

\startlocaldefs
\graphicspath{{figs/}{./}}

\DeclareMathOperator{\R}{\mathbb{R}}

\DeclareMathOperator{\pd}{\partial}

\DeclareMathOperator{\vep}{\varepsilon}

\DeclareMathOperator{\diag}{diag}

\theoremstyle{remark} 
\newtheorem{rem}{Remark}
\newcommand{\proper}{\mathsf}

\newcommand{\pE}{\proper{E}}
\newcommand{\pV}{\proper{V}}

\newcommand{\pN}{\proper{N}}
\newcommand{\mv}[1]{{\boldsymbol{\mathrm{#1}}}}
\newcommand{\trsp}{\ensuremath{\top}}
\newcommand{\md}{\ensuremath{\,\mathrm{d}}}
\newcommand{\scal}[2]{\left\langle {#1},\,{#2} \right\rangle}
\newcommand{\E}{{\bf \proper{E}}}
\newcommand{\V}{{\bf \proper{V}}}
\newcommand{\Vcomp}{{\bf V}}
\newcommand{\vcomp}{V}
\newcommand{\ParFamily}{{\bf \Theta}}
\endlocaldefs

\begin{document}

\begin{frontmatter}
\title{Non-Gaussian Mat\'ern fields with an application to precipitation modeling}
\runtitle{Non-Gaussian Mat\'ern fields}

\author{\fnms{Jonas} \snm{Wallin$^1$}\thanksref{t1}\ead[label=e1]{wallin@maths.lth.se}}
\address{Mathematical Statistics\\
Centre for Mathematical Sciences\\ 
Lund University, Box 118\\ 
SE-22100 Lund, Sweden \\
E-mail:\ \printead*{e1}}
\and
\author{\fnms{David} \snm{Bolin$^2$}\corref{}\ead[label=e2]{david.bolin@math.umu.se}}
\address{Dept of Mathematics\\
and Mathematical Statistics\\
Umeå University\\
SE 90187 Umeå, Sweden\\
E-mail:\ \printead*{e2}}
\affiliation{$^1$Lund University \and $^2$Umeå University}
\runauthor{J. Wallin and D. Bolin}
\thankstext{t1}{The author has been supported by the Swedish Research Council Grant 2008-5382.}
\begin{abstract}
: The recently proposed non-Gaussian Mat\'{e}rn random field models, generated through Stochastic Partial differential equations (SPDEs), are extended by considering the class of Generalized Hyperbolic processes as noise forcings. The models are also extended to the standard geostatistical setting where irregularly spaced observations are modeled using measurement errors and covariates. A maximum likelihood estimation technique based on the Monte Carlo Expectation Maximization (MCEM) algorithm is presented, and it is shown how the model can be used to do predictions at unobserved locations. Finally, an application to precipitation data over the United States for two month in 1997 is presented, and the performance of the non-Gaussian models is compared with standard Gaussian and transformed Gaussian models through cross-validation.
\end{abstract}
\begin{keyword}[class=AMS]
\kwd{62H12,62H11 ,62P12 }
\end{keyword}
\begin{keyword}
\kwd{Mat\'{e}rn covariances}
\kwd{SPDE}
\kwd{Markov random fields}
\kwd{Laplace}
\kwd{Normal inverse Gaussian}
\kwd{MCEM algorithm}
\end{keyword}
\end{frontmatter}

\section{Introduction}
Latent Gaussian models are at the heart of modern spatial statistics. The prime reasons for this are that they are both theoretically and practically easy to work with; there exists a well-developed theory for likelihood-based estimation of parameters and the important problem of spatial reconstruction is easily solved using the standard kriging prediction which is optimal for Gaussian models. For non-Gaussian datasets, the standard approach is to try to find some non-linear transformation that enables the use of Gaussian models. This approach is commonly referred to as trans-Gaussian Kriging  \citep{cressie1993statistics} and common transformations include the square root transform,  \citep{cressie1993statistics, huerta2004spatiotemporal, berrocal2010spatio,sahu2005bayesian} and the log transform \citep{cressie1993statistics, cameletti2011spatio,bolin09b}. An effect of using such transforms is that these induce a certain dependence structure between the mean and the covariance for the data in the untransformed scale. This dependence is often not unreasonable for real data, and it has even been used to generate covariance structures \citep{azais2011transformed}. However, as the models grows more complex, for example by introducing non-stationary covariance functions, spatially varying measurement errors, or covariates, the effects of the transformation methods become less transparent and more stale.
In these situations, one would like to use latent non-Gaussian models without resorting to transformation and the aim of this work is to develop such models.

We state three goals: First, we want to find a class of non-Gaussian models that share some of the desirable properties of the Gaussian models while allowing for heavier tails and asymmetry in the data. Secondly, we want to provide tools for fitting these models to real data, assuming a latent structure with covariates and measurement noise. Finally, we want to provide tools for using the models for spatial reconstruction.

We will extend the work of \cite{bolin11}, where non-Gaussian models with Mat\'ern covariances \citep{matern60} formulated as stochastic partial differential equations (SPDEs) driven by non-Gaussian noise were investigated. The work consisted of providing an existence result for such SPDEs, and in some detail study parameter estimation of SPDEs driven by generalized asymmetric Laplace (GAL) noise. Although this is a good starting point for providing the tools we seek, there are some major issues that have to be resolved in order to use those methods for real applications: The estimation procedure proposed in \cite{bolin11} was based on using the  Expectation Maximization (EM) algorithm, and it works well as long as there is no measurement noise and all nodes in the field are observed. Unfortunately, these requirements are too restrictive for practical applications. However, we will show that these requirements can be avoided, utilizing an  Monte-Carlo Expectation Maximization (MCEM) algorithm, and extend the estimation technique to a larger class of non-Gaussian models.  

The structure of the paper is as follows. In Section~\ref{sec:spde-models}, a brief overview of the methodology used for representing the SPDE models is given. This section also introduces the class of models that is considered in this work, namely SPDE models driven by either GAL noise or Normal inverse Gaussian (NIG) noise and we argue that these two cases are the only relevant cases to consider in the class of generalized hyperbolic distributions for non regular sampled observations.  In section~\ref{sec:modelExt}, we introduce the full hierarchical model that can be used to model spatially irregular observations with covariates and measurement error. In Section~\ref{sec:estimation}, a parameter estimation procedure based on the MCEM is derived. Section ~\ref{sec:Prediction} shows how to do spatial prediction and kriging variance estimation using these models. Section \ref{sec:applications} contains an application of these models to a real dataset consisting of monthly precipitation measurements in the US, and results of the non-Gaussian models are compared with results obtained using standard Gaussian models as well as transformed Gaussian models. Finally, Section~\ref{sec:conclusions} contains some concluding remarks and ideas for future work.

\section{Non-Gaussian SPDE-based models}\label{sec:spde-models}
The Gaussian Mat\'ern fields are perhaps the most widely used models in spatial statistics. These are stationary and isotropic Gaussian fields with a covariance function on the form
\begin{equation}\label{eq:matern}
C(\mv{h}) = \frac{2^{1-\nu}\phi^2}{(4\pi)^{\frac{d}{2}}\Gamma(\nu + \frac{d}{2})\kappa^{2\nu}}(\kappa\|\mv{h}\|)^{\nu}K_{\nu}(\kappa\|\mv{h}\|), \quad \mv{h} \in \R^d, \nu>0,
\end{equation}
where $d$ is the dimension of the domain, $\nu$ is a shape parameter, $\kappa^2$ a scale parameter, $\phi^2$ a variance parameter, and $K_{\nu}$ is a modified Bessel function of the second kind. Since the Mat\'ern-type spatial structure has proven so useful in practice, we want to construct models with this type of spatial structure but with non-Gaussian marginal distributions. In order to do this, we use the fact that a Mat\'ern field $X(\mv{s})$ can be viewed as a solution to the SPDE
\begin{equation}\label{eq:spde}
(\kappa^2-\Delta)^{\frac{\alpha}{2}}X = \dot{M},
\end{equation}
where $\Delta = \sum_{i=1}^d \frac{\pd^2}{\pd\mv{s}_i^2}$ is the Laplacian, and $\alpha = \nu + d/2$ \citep{whittle63}. The Gaussian Mat\'ern fields are recovered by choosing $\dot{M}$ as Gaussian white noise scaled by a variance parameter $\phi$, and the mathematical details of this construction in the case when $\dot{M}$ is non-Gaussian are given in \cite{bolin11}. 

To use these models in practice, we need a method for producing efficient representations of their solutions. One such method is the Hilbert space approximation technique by \cite{lindgren10} which was extended by \cite{bolin11} to the non-Gaussian case when $M(\mv{s})$ is a type G L\'evy process. 

Recall that a L\'{e}vy process is of type G if its increments can be represented as a Gaussian variance mixture $V^{1/2}Z$ where $Z$ is a standard Gaussian variable and $V$ is a non-negative infinitely divisible random variable.  \cite{rosinski91} showed that every type G L\'{e}vy process of can be represented as a series expansion, and for a compact domain $D\in \R^d$ it can be written as $M(\mv{s}) = \sum_{k=1}^{\infty}Z_k g(\gamma_k)^{\frac{1}{2}}\mathbb{I}(\mv{s} \geq \mv{s}_k)$, where the function $g$ is the generalized inverse of the tail L\'{e}vy measure for $V$, $Z_k$ are iid $\pN(0,1)$ random variables, $\gamma_i$ are iid standard exponential random variables, $\mv{s}_k$ are iid uniform random variables on $D$, and 
\begin{equation*}
\mathbb{I}(\mv{s}\geq \mv{s}_k) = \begin{cases}1 & \mbox{if $s_i\geq s_{k,i}$ for all $i \leq d$},\\
0 & \mbox{otherwise.}\end{cases}
\end{equation*} 
Since $V$ is infinitely divisible, there exists a non-decreasing L\'evy process $V(\mathbf s)$ with increments distributed the same as $V$. This process has the series representation $V(\mv{s}) = \sum_{k=1}^{\infty}g(\gamma_k)^{\frac{1}{2}}\mathbb{I}(\mv{s} \geq \mv{s}_k)$.

In the following sections, we briefly describe the Hilbert space approximation technique for the case when $M$ is a type G process, and then introduce a subclass of the type G process that are suitable for the model \eqref{eq:spde}.


\subsection{Hilbert space approximations}
\label{sec:Hilbert}
Assume that $M$ in \eqref{eq:spde} is a type G L\'evy process. The starting point for the Hilbert space approximation method is to consider the stochastic weak formulation of the SPDE,
\begin{equation}\label{eq:weak1}
(\kappa^2 - \Delta)^{\frac{\alpha}{2}}X(\psi) = \dot{M}(\psi),
\end{equation}
where $\psi$ is in some appropriate space of test functions. A finite element approximation of the solution $X$ is then obtained by representing it as a finite basis expansion $X(\mv{s})=\sum_{i=1}^n w_i \varphi_i(\mv{s})$, where $\{\varphi_i\}$ is a set of predetermined basis functions and the stochastic weights are calculated by requiring \eqref{eq:weak1} to hold for only a specific set of test functions $\{\psi_i,i=1,\ldots,n\}$. 
By assuming that $\{\psi_i\} = \{\varphi_i\}$, one obtains a method which is usually referred to as the Galerkin method and this gives an expression for the distribution of the stochastic weights conditionally on the variance process,
\begin{equation}\label{eq:lwodd}
\mv{w} | V \sim \pN\left(\mv{K}_{\alpha}^{-1}\mv{m},\mv{K}_{\alpha}^{-1}\mv{\Sigma}\mv{K}_{\alpha}^{-1}\right).
\end{equation} 
Here $\mv{K}_{\alpha} = \mv{C}(\mv{C}^{-1}\mv{K})^{\alpha/2}$ and the matrices $\mv{K}$, $\mv{C}$, and $\mv{\Sigma}$ have elements given by
$C_{ij} = \scal{\varphi_i}{\varphi_j}$, $K_{ij} = \kappa^2\scal{\varphi_i}{\varphi_j} + \scal{\nabla\varphi_i}{\nabla\varphi_j}$, ${\Sigma_{ij} = \int \varphi_i(\mv{s})\varphi_j(\mv{s}) V(\md \mv{s})}$, and $m_i =  \int \varphi_i(\mv{s}) V(\md \mv{s})$.

In order to get a practically useful representation, we need to be able to evaluate the integrals $\Sigma_{ij}$ and $m_i$ efficiently. Whether this is possible or not depends on the basis $\{\varphi_i\}$ and the variance process $V(\mv{s})$. For the purpose of this work we choose to work with piecewise linear, compactly supported, finite element bases induced by triangulations of the domain of interest. For bases of this type, a mass-lumping procedure gives that $m_i = \vcomp_i$ and $\mv{\Sigma} = \diag(\vcomp_1,\vcomp_2,\ldots,\vcomp_n)$, where 
\begin{equation}\label{eq:variancepart}
\vcomp_i = \int_{h_i} V(\md \mv{s})
\end{equation}
and $h_i$ is the area associated with $\varphi_i(\mv{s})$. For further details, see \cite{bolin11} and \cite{lindgren10}.

\subsection{The generalised hyperbolic processes}
The most well known subclass of the type G L\'evy process is the class of generalised Hyperbolic processes generated by the Generalized Hyperbolic (GH) distribution \citep[see][]{barndorff1978hyperbolic,eberlein2004generalized}. The GH distribution covers a wide range of distributions including the NIG distribution, the Normal inverse Gamma distribution, the GAL distribution, and the $t$-distribution. 

The GH distribution has five parameters $\sigma,\nu\in  \R^+$, $\delta,\mu,\tau\in \R$, and a density function 
\begin{equation}
f(x)=c_1
\left(\frac{\sqrt{ \nu \sigma^2   + (x-\mu)^2 }}{c_2} \right)^{\tau - 1/2} 
K_{\tau - 1/2} \left(c_2 \sqrt{ \nu \sigma^2   + (x-\mu)^2 }\right),
\end{equation} 
where $c_1^{-1}=\left( \frac{\nu \sigma^3}2 \right)^{\tau/2} \sqrt{2\pi} K_{\tau}(\sqrt{2\nu\sigma})$ and $c_2 = \sqrt{\frac1{\sigma^2}(2 + \frac{\mu^2}{\sigma^2})}$. A $GH$ r.v.~$X$ can be represented as
\begin{equation}
\label{eq:NormalMix}
X= \delta + \mu V + \sigma \sqrt{V} Z ,
\end{equation}
where $V$ is a generalized inverse Gaussian r.v.~$V \sim GIG(\tau,\nu^2,2)$ and $Z \sim N(0,1)$. The GIG distribution has the density function
\begin{equation}
\label{eq:GIGdens}
f(x) = \frac{\left ({a/b}\right )^{{p}/{2}}}{2 K_p\left (\sqrt{ab}\right )} x^{p-1}
e^{-\frac{ax + b/x }{2} }.
\end{equation}
where the parameters satisfy $a > 0, b \geq 0$ if $p >0$, $a > 0 , b >0$ if $p =0$, and $a \geq 0 , b> 0$ if $p <0$. Two special cases of the GIG distribution are the inverse Gaussian (IG) distribution which is obtained for $p=-1/2$ and the Gamma distribution which is obtained for $b=0$. We denote the gamma distribution by $\Gamma(a,b)$ and the inverse Gaussian distribution by $IG(a,b)$. For more details of the GIG distribution see \cite{Jorgensen}.

A property of the GH distribution which is important for likelihood-based parameter estimation is that the variance component $V$ is GIG distributed also conditionally on $X$. However, integrals of the variance process $V(\mv{s})$ of a GH process will in general not have known parametric distributions, and thus the random variable $\vcomp_i$ in equation (\ref{eq:variancepart}) will therefore not have known parametric distributions in general. Without this property we are not able to derive likelihood-based parameter estimation procedures, nor make spatial predictions, for the models in this work. 

The random variables $\vcomp_i$ would have known parameteric distributions if the variance process belonged to a class of distributions that is closed under convolution. There are only two special cases of the GH distribution for which the variance components are closed under convolution \citep{PodWal12}. The first special case is the GAL distribution, in finance is known as the variance gamma distribution, which was studied in the context of the SPDE models in \cite{bolin11}, and the second is the NIG distribution. Thus, from now on, we focus on the SPDE model \eqref{eq:spde} driven by either GAL noise or NIG noise.

\begin{rem}
If we would work on regular lattices, there are certain distributions in the GH family, such as the $t$-distribution, where one could imagine fixing the distributions such that the variance process has known distributions for the lattice points; however, having to work on regular lattices is a too strong restriction for us to consider such models any further. Also, even in situations where one has data on a regular lattice and only is interested in predictions to that same lattice, it is not clear what the corresponding continuous model would be if a model of this kind would be used. 
\end{rem}

For the purpose of this work, the most important thing to know about the GAL and NIG distributions is how they affect the Hilbert space approximation procedure. 
For both distributions, we get that $\mv{m}$ and $\mv{\Sigma}$ in the Hilbert space approximation \eqref{eq:lwodd} can be written as $m_i = \gamma\tau h_i + \mu \vcomp_i$, and $\Sigma = \diag(\vcomp_1,\ldots, \vcomp_n)$ respectively. For the GAL distribution $\vcomp(s)$ is a gamma process, and the variance components $\vcomp_i$ are therefore gamma distributed, $\vcomp_i \sim \Gamma(h_i\tau,1)$. For the NIG distribution the $\vcomp(s)$ is a Inverse Gaussian (IG) process, and the variance components are therefore IG distributed, $\vcomp_i \sim IG(\nu^2 h_i,2)$.

\section{Model extensions, covariates, and measurement noise}
\label{sec:modelExt}

To use the models discussed above for real data, we assume a hierarchical model structure. The field of interest, $X(\mv{s})$, is modelled using one of the SPDE models, with observations, $y_1,\ldots, y_N$,  at locations $\mv{s}_1,\ldots, \mv{s}_N$. In practice, these observations are often affected by measurement noise, and we thus need to include this in the model. Furthermore, we allow covariates for the mean value of the field by assuming that $X(\mv{s})$ is on the form
\begin{equation}
X(\mv{s}) = \sum_{i=1}^{n_x} B_{i}(\mv{s})\beta_{i} + \xi(\mv{s}),
\end{equation}
where $\xi(\mv{s})$ is a SPDE field and $\{B_{1},\ldots, B_{n_x}\}$ are known covariates, note that $\xi(\mv{s})$ not necessarily has zero mean in the non-Gaussian case. 
Using the representation \eqref{eq:lwodd} for $\xi(\mv{s})$, where the noise process is on the form of \eqref{eq:NormalMix}, we obtain the following hierarchical model, expressed in terms of the stochastic weights $\mv{w}$ for the basis expansion of $\xi(\mv{s})$ 
\begin{equation}
\label{eq:hierar_old}
\begin{aligned}
\mv{y} &= \mv{B} \mv{\beta} + \mv{A}\mv{w} + \mv{\vep},\\
\mv{w} &= \mv{K}_{\alpha}^{-1}\left(\tau\mv{a}\gamma + \Vcomp\mu + \sigma\sqrt{\Vcomp}\mv{Z}\right).
\end{aligned}
\end{equation}
Here $\mv{A}$ is the observation matrix with elements $A_{ij} = \varphi_i(\mv{s}_j)$ linking the measurements to the latent field, $\mv{B}$ is a matrix containing the covariates $\{B_{i}\}$ evaluated at the measurement locations, and $\mv{\vep}$ is a vector of iid $N(0,\sigma_{\epsilon}^2)$ variables representing the measurement noise.
The vector $\mv{Z}$ contains iid standard Gaussian variables and the distribution of $\vcomp_i$ is determined by the noise process, specifically  $\vcomp_i \sim \Gamma(\tau h_i,1)$ for GAL noise and $\vcomp_i \sim IG(\nu^2 h_i,2)$ for NIG noise and the $\vcomp_i$ are independent, recall that $h_i = \int \varphi_i(\mv{s})\md \mv{s}$. To recover the latent field $X(\mv{s})$ at the measurement locations, one has to calculate $\mv{X} = \mv{B}\mv{\beta} + \mv{A}\mv{w}$.

For the SPDE representation of the Gaussian Mat\'ern fields it is easy to introduce non-stationarity in the model by allowing the covariance parameters to vary with space. In practice, this is achieved by representing the covariance parameters as regressions on some smooth covariates, e.g.~assuming that $\kappa(\mv{s}) = \exp\left(\sum B_{\kappa,i}(s)\beta_{\kappa,i}\right)$ where $\{B_{\kappa,i}\}$ are known covariates would generate a model with a spatially varying covariance range. In the case of the model above, we have several parameters for the noise process, and it might be of interest to allow for these to vary with space as well, especially in cases when one has covariates that not only affect the mean value of the field. This can be achieved in the same way as for the covariance parameters, by assuming regressions on some smooth covariates. For example, we can replace $\gamma$ and $\mu$ in \eqref{eq:hierar_old} by $\gamma(\mv{s}) = \sum B_{\gamma,i}(\mv{s})\gamma_i$ and $\mu(\mv{s}) = \sum B_{\mu,i}(\mv{s})\mu_i$ respectively, where $\{B_{\gamma,i}\}$ and $\{B_{\mu,i}\}$ are smooth covariates. Adding the covariates to \eqref{eq:hierar_old} generates the following hierarchical model:
\begin{equation}
\label{eq:hierar}
\begin{aligned}
\mv{y} &= \mv{B}\mv{\beta} + \mv{A}\mv{w} + \mv{\vep},\\
\mv{w} &= \mv{K}_{\alpha}^{-1}\left(\tau \mv{B}_{\gamma}\mv{\gamma} +\mv{I}_{\Vcomp}\mv{B}_{\mu} \mv{\mu} + \sigma\sqrt{\Vcomp}\mv{Z}\right),
\end{aligned}
\end{equation}
where $\mv{I_{\Vcomp}} = diag(\vcomp_1,\vcomp_2,\ldots,\vcomp_n)$. The matrices $\mv{ B}_{\gamma}$ and $\mv{ B}_{\mu}$ are respectively given by $\{B_{\gamma,i}\}$ and  $\{B_{\mu,i}\}$ evaluated at the node locations.
This is a highly flexible model; however, one needs to be careful in defining the model so that the parameters are identifiable. One needs to be especially careful if using location covariates for both $\bf X$ ($\mv{B}$) and $\bf w$ ($\mv{B}_\gamma$) since this easily leads to a non-identifiable model unless the covariates are chosen carefully to avoid this issue.

\section{Parameter estimation}\label{sec:estimation}
Fitting the model above to data requires a parameter estimation method. In this section, we discuss how the parameters $\ParFamily = \{ \kappa, \mv{\beta}, \sigma_\epsilon ,\tau,\nu\,\mv{\gamma,\mu},\sigma  \}$ can be estimated through likelihood methods for the NIG and GAL-driven SPDEs. The idea is to modify the EM-algorithm in \cite{bolin11}. The modification needed turns out to be the addition of Monte Carlo simulations to estimate the required expectations. We begin with a brief overview of the MCEM-algorithm and then cover the details needed to implement the procedure for our models.

\subsection{Monte Carlo EM} 
The EM-algorithm \citep{Dempster77} is convenient to use when the data-likelihood is difficult to work with but there exists some latent variables $\{{\bf w} ,\Vcomp\}$ so that the augmented data $\{\bf y, w,\Vcomp\}$ has a simpler likelihood (we utilize the same variable names in this subsection as in the rest of the paper for readability, but the result in this subsection is more general then for the models in this paper). The EM-algorithms uses the augmented likelihood  $\pi({\bf y , w, \Vcomp| \ParFamily})$ instead of the original likelihood $\pi({\bf y | \ParFamily})$, but requires the ability to compute expectations of the augmented likelihood.

The $p$th iteration of the EM-algorithm is done in two steps denoted the E-step and the M-step. In the E-step, one computes the function
\begin{align}
\label{eq:Qorig}
\mathcal{Q}\left(\ParFamily,\ParFamily^{(p)}\right) = \pE_{\Vcomp}\left[\log \pi({\bf y, w,  \Vcomp| \ParFamily}) | {\bf y, \Theta}^{(p)}\right],
\end{align}
and in the M-step, one maximizes $\mathcal{Q}(\ParFamily,\ParFamily^{(p)})$ and obtains the $(p+1)$th iterate $\ParFamily^{(p+1)}$. The new iterate has the property $\pi({\bf y | \Theta^{(p+1)}})\ge \pi({\bf y | \Theta^{(p)}})$ and under quite general conditions the procedure converges to a local maximum of the likelihood \citep{wu1983convergence}.

In certain cases when the E-step cannot be calculated analytically, one can use the MCEM algorithm, introduced in \cite{wei1990monte}. The idea of the MCEM algorithm is to replace $\mathcal{Q}$ in the E-step with
\begin{align}
\label{eq:QMC}
 \mathcal{Q}^{MC} \left(\ParFamily,\ParFamily^{(p)} \right) =  \frac{1}{k} \sum_{i=1}^k  \log \pi({\bf y},  \Vcomp^{(i)}, {\bf w}^{(i)} | {\bf \ParFamily}),
\end{align}
where $\{ {\bf w}^{(i)},\Vcomp^{(i)}\}$ is a sample from the distribution $\pi(\Vcomp,{\bf w}|{\bf y, \ParFamily^{(p)}})$. 
In situations where it is not possible sample from the joint density for a set of variables $\{\bf w,\Vcomp\}$, but the conditional densities are available one can use the Gibbs sampling algorithm. The algorithm generates $k$ samples from the joint density by sampling sequentially ${\bf w}^{(i)}| \Vcomp^{(i-1)}$ then $ \Vcomp^{(i)}|{\bf w}^{(i)}$ for $i=1,\ldots, k$. A downside is that the samples $\{{\bf w}^{(i)},\Vcomp^{(i)}\}_{i=1}^k$  will not be independent and also a starting point ${\bf \Vcomp}^{(0)}$ is required.


\subsection{The E-step}\label{sec:e-step}
For the model \eqref{eq:hierar}, the function $\mathcal{Q}$ in \eqref{eq:Qorig} cannot be calculated analytically, and numerical integration is not feasible for the large dimensions of both $\mv{w}$ and $\mv{\Vcomp}$. We therefore use the Monte Carlo method described above to evaluate the E step. 

Ideally we would simulate from $\pi(\Vcomp,{\bf w}| {\bf y},\ParFamily^{(p)})$ in the MC sampler, but the joint distribution for $\bf \{w,\Vcomp\}$ is not known. However, a key observation is that the conditional distributions $\pi(\Vcomp| {\bf w}, {\bf y},\ParFamily)$ and $\pi(\mv{w}| \Vcomp, {\bf y},\ParFamily)$ are known, so we can use a Gibbs sampler to sample from the joint density. 

Note that $\pi(\mv{w}| \Vcomp, {\bf y},\ParFamily)\propto \pi({\bf y}|{\bf w},\Vcomp,\ParFamily)\pi({\bf w}|\Vcomp,\ParFamily)$ where, by construction,
 $\{\mv{y}|{\bf w},\Vcomp,{\bf \Theta}\}$ and $\{{\bf w}|\Vcomp,\ParFamily\}$ are Gaussian, and $\{\mv{w}| \Vcomp, {\bf y},\ParFamily\}$ is therefore also Gaussian.  The explicit form  of $\pi({\bf w}|\{\Vcomp,{\bf y, \ParFamily}\})$ is $N({\bf \hat{m}}, {\bf \hat{Q}}^{-1})$ where 
\begin{alignat*}{2}
\hat{\mv{m}}  &=\hat{\mv{Q}}^{-1}\left(\mv{Q}\mv{m}  + \frac1{\sigma_{\epsilon}^2}\mv{A}^{\trsp} (\mv{y}-\mv{B}\mv{\beta})\right)   
 &\, ,\hat{\mv{Q}} &= \mv{Q} + \frac1{\sigma_{\epsilon}^2}\mv{A}^{\trsp}\mv{A}, \\
 \mv{m}&=\mv{K}^{-1}_{\alpha}(\mv{B}_\gamma \mv{\gamma} + \mv{I}_{\mv{\Vcomp}}\mv{B}_\mu \mv{\mu}) \,, &\mv{Q}&=\frac1{\sigma^2} \mv{K}_{\alpha}\mv{I}^{-1}_{\mv{\Vcomp}}\mv{K}_{\alpha}.
\end{alignat*}

The density of $\{\Vcomp|{\bf w},{\bf y,  \ParFamily} \}$ is proportional to
\begin{equation*}
\pi({\bf y}|{\bf w},\Vcomp, \ParFamily)\pi({\bf w}|\Vcomp, \ParFamily)\pi(\Vcomp| \ParFamily) \propto \pi({\bf w}|\Vcomp, \ParFamily)\pi(\Vcomp| \ParFamily).
\end{equation*}
For both GAL and NIG processes, $\pi(\Vcomp|{\bf  \ParFamily})$ can be written as $GIG({\bf p,a,b})$ and we therefore get
\begin{align*}
\pi(\Vcomp|{\bf w},{\bf y,  \ParFamily}) \propto &
\left(\prod_j \vcomp_j^{p_j-1} \right)\left(\prod_j \vcomp_j^{1/2} \right)
\exp \Big(-\frac1{2}\left( {\bf 1}^{\trsp}{\bf I}_{\bf \Vcomp}{\bf a}-  {\bf 1}^{\trsp}{\bf I}^{-1}_{ \Vcomp}{\bf b} \right) 
\\
&- \frac1{2\sigma^2}  (\mv{K}_{\alpha}\mv{w} - \mv{B}_{\gamma} \mv{\gamma} - {\bf I}_{\bf \Vcomp}\mv{B}_{\mu} \mv{\mu}  )^{\trsp}  {\bf I}^{-1}_{ \Vcomp}(\mv{K}_{\alpha}\mv{w} - \mv{B}_{\gamma} \mv{\gamma} - {\bf I}_{\bf \Vcomp}\mv{B}_{\mu} \mv{\mu}  ) \Big) 
 \\
 =& \prod_{j}  \vcomp_j^{p_j-1/2} \exp \Big(- \frac1{2}\left( \frac{ ({\bf K_{\alpha}w} - \mv{B}_{\gamma} \mv{\gamma})_j^2}{\sigma^2}  + b_j \right)\vcomp^{-1}_j \\
 &-\frac1{2}\left(\frac{(\mv{B}_{\mu} \mv{\mu} )^2_j}{\sigma^2}  + a_j \right) \vcomp_j  \Big),
\end{align*}
which is a GIG distribution with parameters given in Table~\ref{tab:Vcomp} for the NIG and GAL cases.

\begin{table}
\centering
\begin{tabular}{@{}lcc@{}}
\toprule
 & $GAL$ & $NIG$  \\
\cmidrule(r){1-3}
 ${\bf p} $  & ${\bf h}\tau-1/2$ & ${\bf -1}$ \\ 
$\mv{a} $ & $(\mv{B}_\mu\mv{\mu})^2/\sigma^2 + 2$ & $(\mv{B}_\mu\mv{\mu})^2/\sigma^2 + 2$  \\ 
${\bf b} $& $ (\mv{K}_{\alpha}\mv{w} - \mv{B}_\gamma\mv{\gamma})^2/\sigma^2$ & $(\mv{K}_{\alpha}\mv{w}-\mv{B}_\gamma\mv{\gamma})^2/\sigma^2 + \mv{h} \nu^2$\\
\bottomrule
\end{tabular}
\caption{The distribution of $\{\Vcomp|{\bf w}, \ParFamily\}$, used in the Gibbs sampler, is $GIG({\bf p,a,b})$ with parameters given in the table for the cases of NIG noise and GAL noise. Note that the distribution is independent of $\bf Y$ in both cases.}
\label{tab:Vcomp}
\end{table}

\subsection{The M-step}
\label{sec:Mstep}
To find the updating equations for the parameters, $\mathcal{Q}^{MC}$ should be maximized. The log-likelihood $\log \pi(\mv{y}, \mv{\Vcomp}^{(i)}, \mv{w}^{(i)} | \mv{\ParFamily})$ can be divided into three terms
\begin{equation}\label{eq:full_like}
\log \pi(\mv{y}|\mv{w}^{(i)}, \mv{\Vcomp}^{(i)},\mv{\ParFamily})  +
\log \pi(\mv{w}^{(i)}|\mv{\Vcomp}^{(i)},\mv{\ParFamily}) + 
\log \pi(\Vcomp^{(i)}|\mv{\ParFamily}).
\end{equation}
The first term on the right hand side is a function of  $\mv{\ParFamily}$ only through $\{\mv{\beta}, \sigma_\epsilon\}$, the second term only through $\{\mv{\gamma,\mu},\sigma,\kappa\}$, and the third term only through  
$\{\tau,\nu\}$, together with that the first term is independent of $\mv{\Vcomp}^{(i)}$ enables us to rewrite equation \eqref{eq:full_like} as
\begin{equation}\label{eq:full_like2}
\log \pi(\mv{y}^{(i)}|\mv{w}^{(i)}, \mv{\beta}, \sigma_\epsilon)  +
\log \pi(\mv{w}^{(i)}|\mv{\Vcomp}^{(i)},\mv{\gamma,\mu},\sigma,\kappa) + 
\log \pi(\Vcomp^{(i)}|\tau,\nu).
\end{equation}
Thus, the joint maximization of \eqref{eq:full_like2} for $\mv{\ParFamily}$ can be split into three  separate steps, where maximization over $\{\tau,\nu\}$, $\{\mv{\beta}, \sigma_\epsilon\}$ and $\{\mv{\gamma,\mu},\sigma,\kappa\}$ is preformed independently. 

The part of the log-likelihood depending on $\{\tau,\nu\}$ is
\begin{equation*}
\log \pi(\Vcomp|\tau,\nu^2) = c+  \begin{cases}\tau {\bf h}^{\trsp} \log \Vcomp- \sum_{i=1}^ n  \log \Gamma( \tau h_i)  & \mbox{for GAL,}\\
n\log(\nu) + \sqrt{2}\mv{1}^{\trsp} \mv{h}^{1/2}\nu -  \frac1{2} {\bf h^{\trsp} \Vcomp}^{-1} \nu^2 & \mbox{for NIG,}
\end{cases}
\end{equation*}
where $c$ is a constant.
The maxima with respect to these parameters are given in Table \ref{tab:VcompLike}.
Note that this is the only part of the M step where the estimation for the NIG and GAL models differ. For the NIG model, the updating equation for $\nu^2$ is given analytically whereas one has to do numerical optimization to update $\tau$ in the GAL model.
\begin{table}
\centering
\begin{tabular}{@{}lcc@{}}
\toprule
 & $GAL$ & $NIG$  \\
\cmidrule(r){1-3}
 $\tau $& $\max_{\tau} \, \frac{\tau}{k} \mv{h}^{\trsp} \left(\sum_{i=1}^k\log \vcomp^{(i)} \right)- \sum_{j=1}^ n  \log \Gamma( \tau h_j) $ &  $-1/2$ \\
 $\nu^2$ & $0$ & $\left(\frac{\mv{1}^{\trsp}\mv{h}^{1/2}  + \sqrt{(\mv{1}^{\trsp} \mv{h}^{1/2})^2 + 2n\mv{h}^{\trsp} \bar{\mv{\Vcomp}}^{-1}}}{\sqrt{2} \mv{h}^{\trsp}\bar{\mv{\Vcomp}}^{-1}}\right)^2$ \\ 
\bottomrule
\end{tabular}
\caption{The parameter values that maximizes the function $\log \pi(\Vcomp|\tau,\nu^2)$ for the cases of GAL and NIG noise. Here $\bar{\Vcomp}^{-1} = \frac1k \sum_{i=1}^k (\Vcomp^{(i)})^{-1}$.} 
\label{tab:VcompLike}
\end{table}

To update $\{\mv{\beta},\sigma_{\epsilon}\}$, one should maximize $\sum_i \log \pi (\mv{y}, \mv{w}^{(i)} | \mv{\beta}, \sigma_{\epsilon})$,  where
\begin{align*}
\log \pi (\mv{y}, \mv{w} | \mv{\beta}, \sigma_{\epsilon}) =&
 -\frac1{2\sigma_\epsilon^2} \left(\mv{y} -\mv{Aw} - \mv{B}\mv{\beta} \right)^{\trsp}  \left(\mv{y} -\mv{Aw} - \mv{B}\mv{\beta} \right)\\
&-n\log(\sigma_\epsilon) - \frac{n}{2}\log(2\pi).
\end{align*}
The function is maximized by $\mv{\beta} = (\mv{B}^{\trsp} \mv{B})^{-1} \mv{b}_x$ and $\sigma_\epsilon^2=(H_x - \mv{B}_x^{\trsp} \mv{\beta})$ where 
\begin{align*}
\mv{b}_x &= \frac1k \sum_{i=1}^k  (\mv{y}-\mv{Aw}^{(i)})^{\trsp} \mv{B}, & 
 H_x &= \frac1k \sum_{i=1}^k (\mv{y}-\mv{Aw}^{(i)})^{\trsp} (\mv{y}-\mv{Aw}^{(i)}).
\end{align*}

In the third step, we find the maximum of the likelihood for $\{ \mv{\gamma,\mu},\sigma,\kappa \}$, which only requires maximization of $\sum_i \log \pi(\mv{w}^{(i)}|\mv{\Vcomp}^{(i)},\mv{\gamma,\mu},\sigma,\kappa)$. The estimation needs to be done jointly for these parameters, and in general there is no closed form solution. However, it is possible to split this estimation step into two conditional maximization steps as described in \cite{bolin11}. An alternative is to use the fact that we can calculate the maximum of the function for a fixed $\kappa$ by maximizing $\sum_i \log \pi(\mv{w}^{(i)}|\mv{\Vcomp}^{(i)},\mv{\gamma,\mu},\sigma,\kappa)$
over $\{ \mv{\gamma,\mu},\sigma\}$. For a fixed $\kappa$, this function is maximized by
\begin{alignat*}{2}
\begin{bmatrix} 
 \mv{\mu} \\
 \mv{\gamma} 
\end{bmatrix} & = {\bf Q}^{-1}_{par} {\bf b}, & \quad
\sigma^2=
\frac1{n}\left(H - {\bf b}^{\trsp}
\begin{bmatrix} 
 \boldsymbol \mu \\
  \boldsymbol  \gamma
\end{bmatrix}\right),
\end{alignat*}
where 
\begin{align*}
{\bf Q}_{par} &= \frac1k \sum_{i=1}^k 
\begin{bmatrix}
{\bf B_\mu^{\trsp} I}_{\Vcomp^{(i)}}{\mv B}_\mu & {\bf B_\mu^{\trsp} B_\gamma} \\
{\bf B_\mu B_\gamma^{\trsp}}  & {\bf B_\gamma^{\trsp} I}_{\Vcomp^{(i)}}^{-1}{\bf B_\gamma}
 \end{bmatrix}, \,
 {\bf b} =  \frac1k \sum_{i=1}^k \begin{bmatrix}  ({\bf K_{\alpha}w}^{(i)})^{\trsp} {\bf B_\mu} \\ ({\bf K_{\alpha}w}^{(i)})^{\trsp} {\bf I}_{\Vcomp^{(i)}}^{-1} {\bf B_\gamma}  \end{bmatrix}, \\
H &=\frac1k\sum_{i=1}^k {\bf (K_{\alpha}w}^{(i)})^{\trsp} {\bf I}_{\Vcomp^{(i)}}^{-1} {\bf K_{\alpha}w}^{(i)}.
\end{align*}
Inserting these expressions for $\{ \mv{\gamma,\mu},\sigma\}$ into $\sum_i \log \pi(\mv{w}^{(i)}|\mv{\Vcomp}^{(i)},\mv{\gamma,\mu},\sigma,\kappa)$ yields an equation which is maximized numerically with respect to $\kappa$ to find the new values for $\{ \mv{\gamma,\mu},\sigma,\kappa \}$. For $\alpha=2$, this equation is given by
\begin{equation*}
-\log(|\mv{K}_{\alpha}|) + \frac{n}{2}\log(H-\mv{b}^{\trsp}\mv{Q}_{par}\mv{b})
\end{equation*}
and similar, though more involved expressions can be found for other even values of $\alpha$ since $\bf K_{\alpha}$ can be written as a matrix polynomial in these cases. 

A potential problem with the MCEM algorithm is that it could require a lot of memory if all values of $\{\Vcomp^{(i)},{\bf w}^{(i)}\}$ for $i=1,\ldots,k$ needed to be stored in order to evaluate the M step. However, as seen above, we only need to store a number of sufficient statistics in order to evaluate the M step. For $\alpha = 2$, these are given by   
\begin{align*}
&\sum_{i=1}^k(\mv{Cw}^{(i)})^{\trsp}\mv{I}_{\Vcomp^{(i)}}^{-1} \mv{Cw}^{(i)}, & 
&\sum_{i=1}^k(\mv{Cw}^{(i)})^{\trsp} \mv{I}_{ \Vcomp^{(i)}}^{-1} {\bf  B}_\gamma, &
&\sum_{i=1}^k(\mv{Gw}^{(i)})^{\trsp} {\bf  B}_\mu,
 \\
&\sum_{i=1}^k(\mv{Cw}^{(i)})^{\trsp} \mv{I}_{\Vcomp^{(i)}}^{-1} \mv{Gw}^{(i)}, &
&\sum_{i=1}^k(\mv{Gw}^{(i)})^{\trsp} \mv{I}_{ \Vcomp^{(i)}}^{-1} {\bf  B}_\gamma, & 
&\sum_{i=1}^k(\mv{Cw}^{(i)})^{\trsp} {\bf  B}_\mu, 
\\
&\sum_{i=1}^k(\mv{Gw}^{(i)})^{\trsp} \mv{I}_{\Vcomp^{(i)}}^{-1} \mv{Gw}^{(i)}, &
&\sum_{i=1}^k {\bf B_\gamma^{\trsp} I}_{\Vcomp^{(i)}}^{-1}{\bf B_\gamma}, &
&\sum_{i=1}^k {\bf B_\mu^{\trsp} I}_{\Vcomp^{(i)}}{\bf B_\mu}.
\end{align*}
Thus, for $\alpha=2$, we only need to store nine values to evaluate the M step. As $\alpha$ increases number of sufficient statistics required for storage will increase, but for any reasonable value of $\alpha$ the number of sufficient statistics is much smaller than the number of elements in $\{\Vcomp^{(i)},{\bf w}^{(i)}\}$.

\begin{rem}
For the GIG distribution, $\pE(\Vcomp^{-1})$ can be unbounded when $|p|$ is small and $b \rightarrow 0$. This makes the estimation of $\kappa$ and $\gamma$ problematic when $\min \left(| \tau{\bf h} -1/2|\right)$ is small for the GAL model. The same problem exists for the EM algorithm in \cite{bolin11}, and that work gives some suggestions on how to improve the estimation in this situation. 
\end{rem}

\subsection{Rao-Blackwellization}
\label{sec:Pract}
For each MC sample in the $E$-step, a sample ${\bf w}^{(i)}$, from $\pi(\mathbf{w|y,\Theta, \Vcomp})$, is required.
Sampling ${\bf w}^{(i)}$ requires a Cholesky decomposition of $\hat{\mv{Q}}$ which in general has a computational cost of $O(n^{3/2})$ for the SPDE models on $\R^2$, where $n$ is the number of elements in $\mv{w}$. The Cholesky factorization dominates the total computational cost of the $E$-step, which in turn dominates the total computational cost of the MCEM algorithm. Thus, in order to reduce the computational cost of the estimation it is crucial to reduce the number of MC simulations in the $E$-step. 

A common trick that can be used to reduce the number of required MC simulations to achieve a certain variance of the estimator is to note that for any function $h$ and any two random variables $X$ and $Y$, one has that $\E\E [[h(X)|Y]] = \E [h(X)]$ and $\V [\E [h(X)|Y]] \leq \V[h(X)]$. When this is used in estimation, it is usually referred to as Rao-Blackwellization \citep[see][]{robert2004monte} due to its association with the Rao-Blackwell Theorem \citep[see][]{ferguson1967mathematical}.

To apply Rao-Blackwellization to $\mathcal{Q}^{MC}$, we note that $\mathcal{Q}\left(\ParFamily,\ParFamily^{(p)}\right)$ can be written as
$\E \left[\E \left[ \log \pi({\bf y,  \Vcomp,w | \ParFamily}) |\star\right]|{\bf y}, \ParFamily^{(p)}\right]$,
where $\star$ denotes $\{{\bf y,w}, \ParFamily^{(p)}\}$, the inner expectation is taken over $\Vcomp$, and the outer expectation is taken over $\mv{w	}$.
Viewing the log likelihood in equation \eqref{eq:full_like} as a function of $\Vcomp$, one sees that   
\begin{align*}
\pE[\log\pi({\bf y,\Vcomp,w | \ParFamily}) |\star] =&
- \frac1{2\sigma^2} \left( ({\bf K_{\alpha}w} - {\bf B_\gamma \boldsymbol \gamma})^{\trsp}  \E[{\bf I}_{\bf \Vcomp^{-1}}|\star]({\bf K_{\alpha}w}- {\bf B_\gamma \boldsymbol \gamma}  )\right. \\
&\left. +  {\boldsymbol \mu^{\trsp} \bf B_\mu^{\trsp}}\E[{\bf I}_{\bf \Vcomp}|\star]{\bf B_\mu \boldsymbol \mu} \right) - \E [\log \pi(\Vcomp|\tau,\nu^2)|\star]
\end{align*}
up to an additive constant, as a function of $\Vcomp$, where the last term is
\begin{equation*}
\pE[\log \pi(\Vcomp|\tau,\nu^2)|\star] = c+
\begin{cases}
2^{-1} {\bf h^{\trsp}}\nu^2 \pE[\Vcomp^{-1}|\star],  &\mbox{for NIG noise},\\
\tau {\bf h}^{\trsp} \E[\log \Vcomp|\star],  &\mbox{for GAL noise}.
\end{cases}
\end{equation*}
We therefore have the option to replace $\mathcal{Q}^{MC}$ with
\begin{align*}
 \mathcal{Q}^{RB}({\bf \Theta,\Theta}^{(p)}) =  \frac1k \sum_{i=1}^k\E \left[ \log \pi({\bf Y,  \Vcomp,w}^{(i)} | \ParFamily)   | {\bf Y,w}^{(i)}, \ParFamily^{(p)} \right],
\end{align*} 
which is a Rao-Blackwelliztion of $\mathcal{Q}^{MC}({\bf \Theta,\Theta}^{(p)})$.
Here, the expectations $\pE[\Vcomp|\star]$, $\pE[\Vcomp^{-1}|\star]$, and $\pE[\log\Vcomp|\star]$ can be computed numerically using the following formulas for the expectations of a $GIG(p,a,b)$ random variable $\vcomp$
\begin{equation*}
\begin{split}
\E [\vcomp^{\lambda}] &= (b/a)^{\lambda /2}\frac{ K_{p+\lambda}\left (\sqrt{ab}\right )}{  K_p\left (\sqrt{ab}\right ) },~~\lambda\in \mathbb R\\
\E [\log(\vcomp)] &= 
\log (\sqrt{{a}/{b}}) 
+ 
\frac{\partial}{\partial p} \log K_{p}\left (\sqrt{ab}\right ).
\end{split}
\end{equation*}
The expectation $\E [\log(\vcomp)]$ can be approximated by approximating 
\begin{equation*}
\frac{\partial}{\partial p} \log K_{p}\left (\sqrt{ab}\right ) \approx \left(\log K_{p+\vep}\left(\sqrt{ab}\right) - \log K_{p}\left(\sqrt{ab}\right)\right)/\vep 
\end{equation*}
for some small $\vep>0$.

\section{Prediction}\label{sec:Prediction}
One of the main problems in spatial statistics is prediction of the latent field at locations where there are no observations. The two main characteristics that are reported in such predictions are the mean and variance of the predictive distribution. In this section, we show how to generate these two quantities for predictions, using the models described previously, at a set of locations $\mv{s}_1,\ldots,\mv{s}_p$.

 Let ${\bf A}_p$ be a $p\times n$ observation matrix, constructed the same way as the observation matrix in Section \ref{sec:modelExt}, for the locations $\mv{s}_1,\ldots,\mv{s}_p$. The desired mean values and variances are $\pE [\mv{A}_p \mv{w}|\mv{y},\ParFamily]$ and $ \V [\mv{A}_p \mv{w}|\mv{y},\ParFamily]$ respectively. Since the density of $\mv{w}|\mv{y}$ is not known, the mean and variance cannot be calculated analytically, and we therefore utilize MC methods to approximate the mean as $ \E [\mv{A}_p \mv{w}|\mv{y},\ParFamily] \approx \frac1k \sum_{i=1}^k \mv{A}_p\mv{w}^{(i)}$ and the variance as $\V [\mv{A}_p \mv{w}|\mv{y},\ParFamily] \approx \frac1{k} \sum_{i=1}^k (\mv{A}_p\mv{w}^{(i)} - \E [\mv{A}_p \mv{w} |\mv{y},\ParFamily])^2$,
where ${\bf w}^{(i)}$ is generated using the Gibbs-sampler described in Section \ref{sec:e-step}. 

Rao-Blackwelliztion can again be used to reduce the variance of the MC estimates. For the mean, write
\begin{align*}
 \E [\mv{A}_p \mv{w}|\mv{y},\ParFamily] &=   \int_{\mv{w}} \mv{A}_p\mv{w} \pi(\mv{w}|\mv{y},\ParFamily) \md \mv{w} \\
&=  \int_{\mv{w}} \int_\Vcomp \mv{A}_p \mv{w} \pi(\mv{w}|\Vcomp,\mv{y},\ParFamily)\pi(\Vcomp|\mv{y},\ParFamily) \md \Vcomp\md \mv{w} \\
&=  \int_{\Vcomp} \mv{A}_p\hat{\mv{m}} \pi(\Vcomp|{\bf y},\ParFamily)  d\Vcomp \approx \frac1k \sum_{i=1}^k \mv{ A}_p\hat{\mv{m}}^{(i)},
\end{align*}
which is a Rao-Blackwelliztion of $ \E [\mv{A}_p \mv{w}|\mv{y},\ParFamily]$ where $\hat{\mv{m}}$ is the conditional mean of $\mv{w}$, defined in Section \ref{sec:e-step}. Since the Gibbs-sampler uses $\hat{\mv{m}}^{(i)}$ to simulate  $\mv{w}^{(i)}$, the Rao-Blackwelliztion can be produced from the MC sampler in the estimation step with no extra cost. The Rao-Blackwellization for the variance of the prediction is derived similarly as
\begin{align}\label{eq:RBvar}
\bf \V [\mv{A}_p \mv{w}|\mv{y},\ParFamily] &=
  \int_{\mv{w}} \mv{A}_p\left(\mv{w - \hat{w}}\right)\left(\mv{w - \hat{w}}\right)^{\trsp}\mv{A}_{p}^{\trsp} \pi(\mv{A}_p \mv{w}|\mv{y},\ParFamily) \md \mv{w} \notag\\
&=   \int_{\Vcomp} \mv{A}_p\hat{\mv{Q}}^{-1}\mv{A}_{p}^{\trsp}\pi(\Vcomp|\mv{y},\ParFamily)\md \Vcomp
\approx\frac1k \sum_{i=1}^k {\bf A}_p^{\trsp} \left({\bf \hat{Q}}^{(i)} \right)^{-1} {\bf A}_p.
\end{align}
It would seem as one needs to calculate the inverse of ${\bf \hat{Q}}^{(i)}$, which is computationally expensive, to use Rao-Blackwellization of the variances. However, because of the structure of ${\bf A}_p$, only the elements of the inverse of $ {\bf \hat{Q}}^{(i)}$ that corresponds to the non-zero elements in $ {\bf \hat{Q}}^{(i)}$ are needed to evaluate \eqref{eq:RBvar}. Using the methods in \cite{campbell1995computing}, one can compute these elements at a computational cost of  $O(n^{3/2})$, making Rao-Blackwellization for the variances computationally feasible.

\begin{figure}
\input{figs/RaoBlack1.tex}
\includegraphics[scale=0.5]{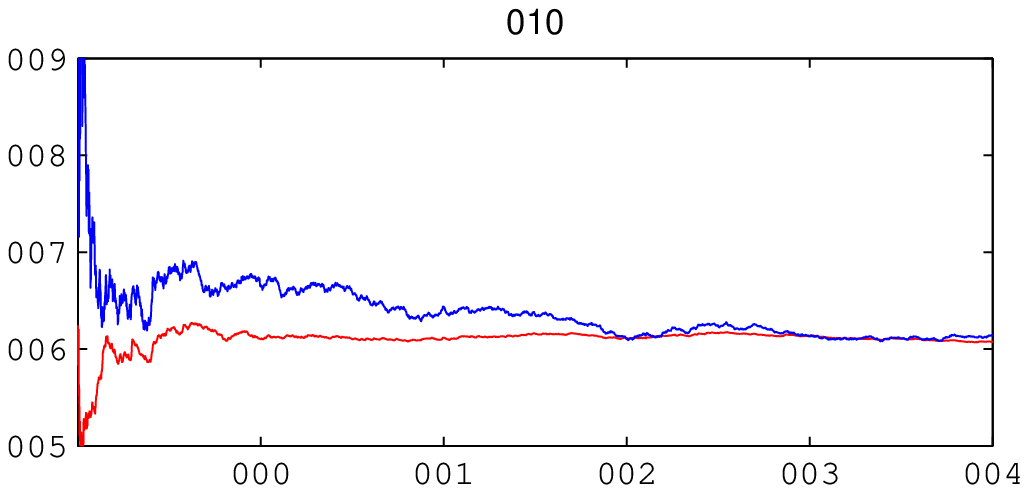}
\input{figs/RaoBlack2.tex}
\includegraphics[scale=0.5]{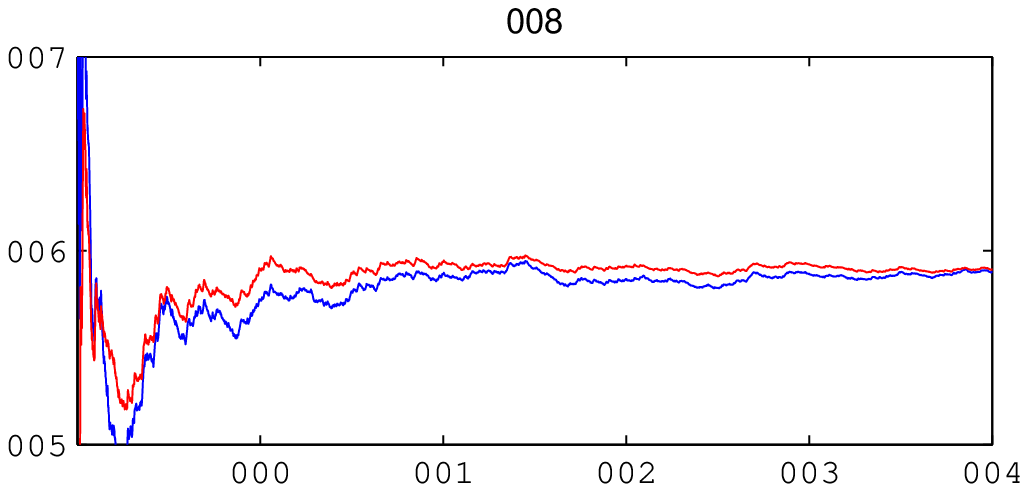}
\caption{The figure shows the convergence of the Rao-Blackwellization estimator (red lines) and the regular Monte-Carlo estimator (blue lines) when estimating the conditional mean of a field at two distinct locations. For the first location (left panel), the Rao-Blackwellization improves the convergence, whereas the Rao-Blackwellization has no noticeable effect for the second location (right panel).}
\label{fig:KrigRao}
\end{figure}

To illustrate the effect of the Rao-Blackwellization, we examine the convergence of the Monte-Carlo estimator and the Rao-Blackwellization for the estimation of two conditional means at two distinct locations, $m_1 =  \E [\mv{A}_1 \mv{w}|\mv{y},\ParFamily]$ and $m_2 = \E [\mv{A}_2 \mv{w}|\mv{y},\ParFamily]$, of the precipitation data used in Section \ref{sec:applications}. The results can be seen in Figure \ref{fig:KrigRao}, the convergence of the estimation of $m_1$ is seen in the left panel and the convergence of the estimation of $m_2$ is seen in the right panel. As seen in the figure, the Rao-Blackwelliztion has a large effect on the convergence for $m_1$ whereas it has no visible effect on the convergence for $m_2$. The reason for this difference is that the largest part of variance of the MC method for $m_1$ comes from ${\bf w}|\Vcomp$ whereas the largest part of variance for $m_2$ comes from the variance of $\Vcomp|{\bf w}$.

\section{An application to precipitation modeling}\label{sec:applications}
One of the most important aspects of geostatistical models is the ability to do spatial infilling of environmental data to produce high-resolution maps of the modeled quantities, and an equally important property is the ability to produce uncertainty estimates for those infilled maps. Such high-resolution maps are used for a number of different purposes, ranging from aiding in getting a better understanding of the earth system to studying the effects of climate change and assessing climate models. In this section, we consider an application of the non-Gaussian latent models of the previous sections to producing high-resolution maps of precipitation. The purpose of this application is twofold. Firstly, it serves as an illustration of the fact that the models we have proposed can be used for analyzing large environmental datasets. Secondly, we want to investigave if anything is gained by using a fully non-Gaussian model compared with the simpler alternative of using Gaussian models for transformed data, which has previously been used for this dataset.

\begin{figure}[t]
\begin{center}
\begin{minipage}[b]{0.4\linewidth}
\centering
January 1997\\
\includegraphics[width=\linewidth,bb=-0 -0 452 198,clip=]{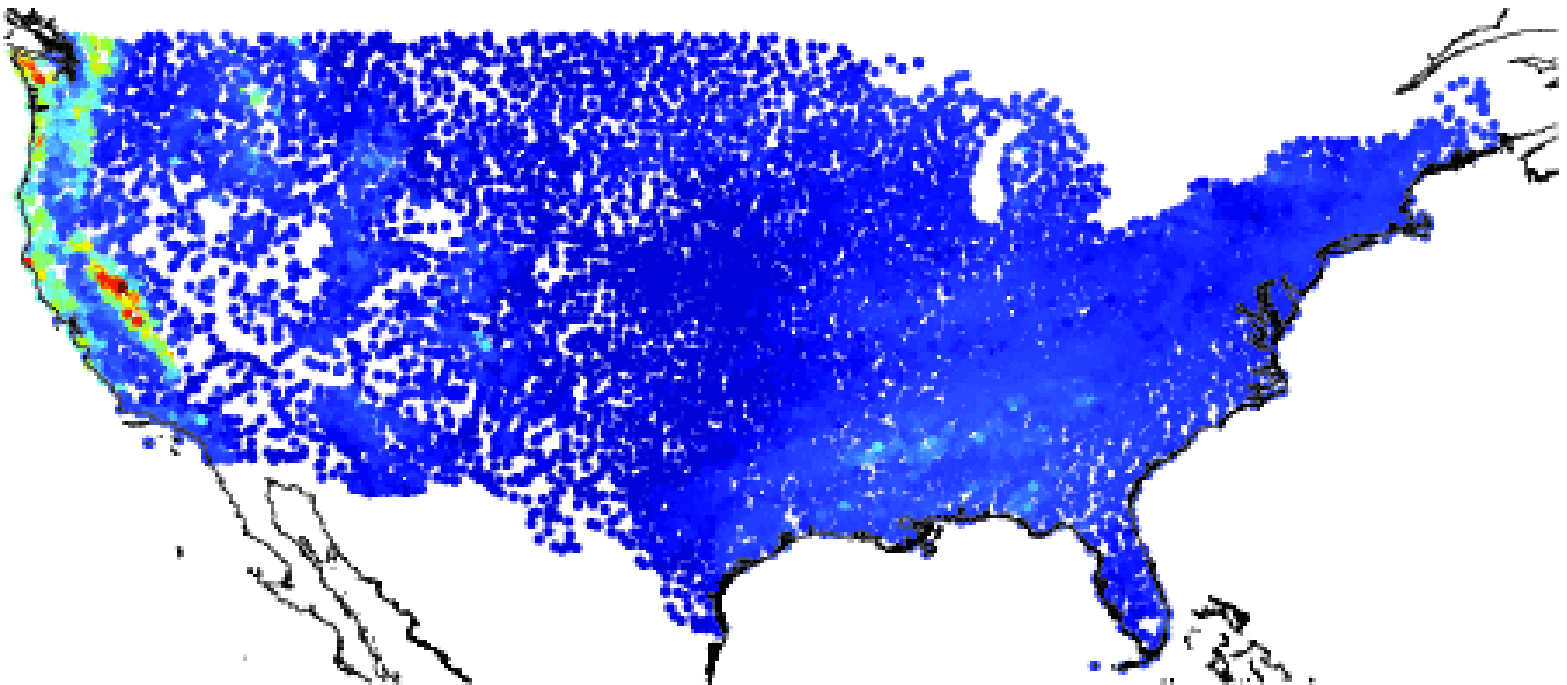}%
\end{minipage}
\begin{minipage}[b]{0.07\linewidth}
\centering
\quad\\
\input{figs/precip_jan_cb.tex}
\includegraphics[height=25mm,bb= 65 2856 102 2978,clip=]{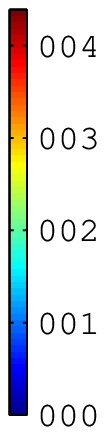}%
\end{minipage}
\begin{minipage}[b]{0.4\linewidth}
\centering
June 1997\\
\includegraphics[width=\linewidth,bb=-0 -0 452 198,clip=]{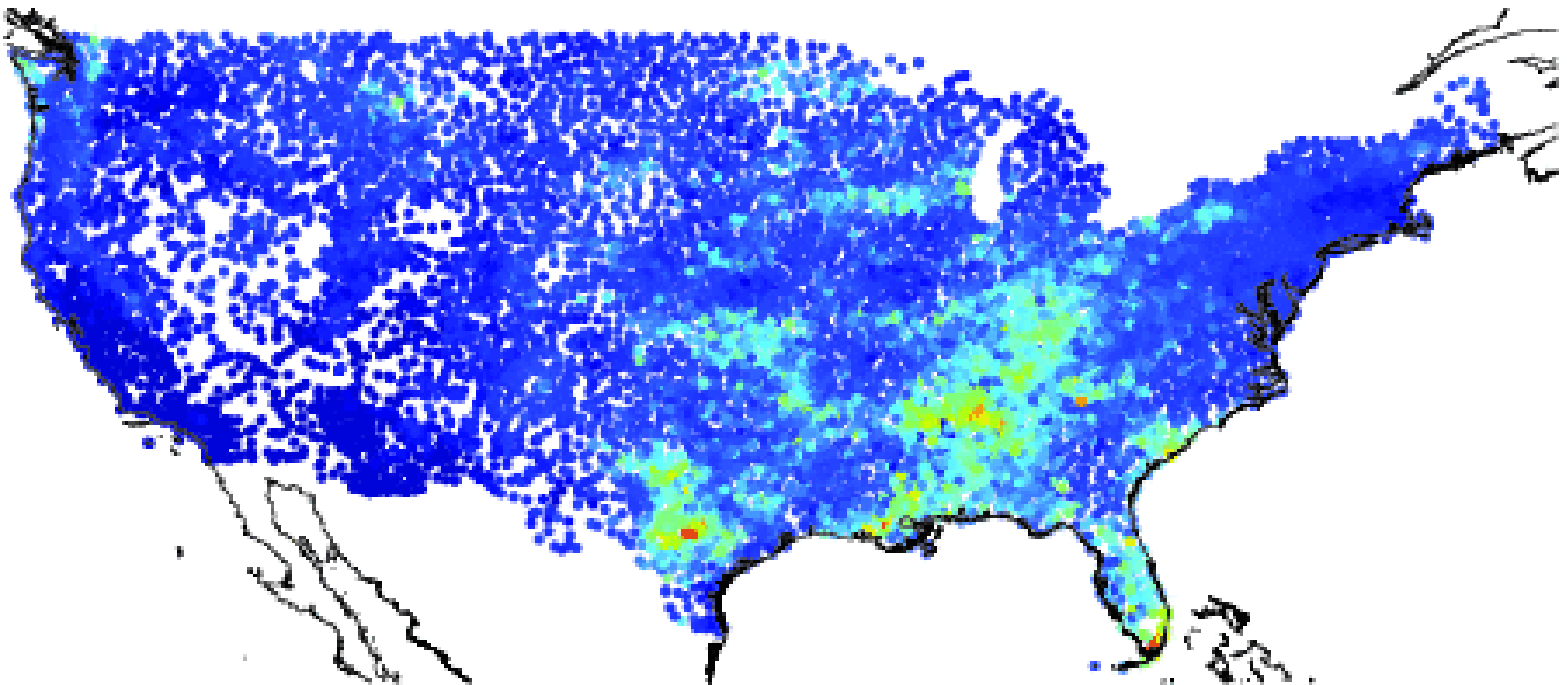}%
\end{minipage}
\begin{minipage}[b]{0.07\linewidth}
\centering
\quad\\
\input{figs/precip_jun_cb.tex}
\includegraphics[height=25mm,bb= 65 2856 102 2978,clip=]{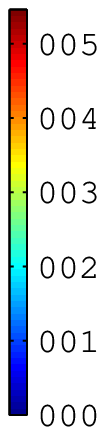}%
\end{minipage}
\end{center}
\vspace{-0.4cm}
\caption{Precipitation data (mm) for January (left) and June (right) 1997.}
\label{fig:precip_data}
\end{figure}

The data we use is available online\footnote{
\texttt{www.image.ucar.edu/Data/US.monthly.met}}
and was created from the data archives of the National Climatic Data Center.
The data is measured at $11918$ unique sites throughout the United States, and has a temporal coverage for the period 1895-1997. The spatial coverage varies over the time period, with fewer stations reporting in the beginning. The actual measurements are reports of total monthly precipitation in mm scale, see Figure \ref{fig:precip_data} for data from January and June 1997. For more details on the data, see \cite{johns2003infilling}. 

\subsection{Models}
We will use four different models to analyze the data. The first three are models for the data in the original scale and the fourth is a Gaussian model for square-root transformed data. 

\begin{figure}[t]
\begin{center}
\includegraphics[width=1.1\linewidth,height=4cm,bb=240 40 1200 900,clip=]{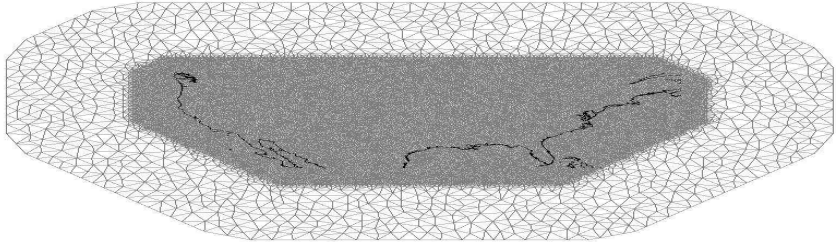}%
\end{center}
\vspace{-0.4cm}
\caption{The triangulation with $33235$ nodes used for the Hilbert space approximations. The coastlines are shown in black. The reason for extending the triangulation outside the region of interest with large triangles is to reduce boundary effects in the SPDE representation.}
\label{fig:triangulation}
\end{figure}

For the first three models, we assume that the measurements, $y_i$, are generated as $y_i = X(\mv{s}_i) + \vep_i$, where $\vep$ is Gaussian measurement noise with variance $\sigma_{\vep}^2$ and $X(\mv{s}) = \beta + \xi(\mv{s})$ is the latent precipitation field which we model as a stationary Mat\'ern field. We fix the shape parameter $\alpha$ of the Mat\'ern covariance function at two, but estimate the other parameters from the data. The SPDE representation is used for $\xi$, and the basis for the Hilbert space approximation is chosen as the basis of piecewise linear basis functions induced by the triangulation in Figure \ref{fig:triangulation}. The three different models are obtained by choosing the forcing noise in the SPDE as either Gaussian noise, GAL noise, or NIG noise. 

For the final model, the precipitation measurements are modeled as
$\sqrt{y_i} = X(\mv{s}_i) + \vep_i$, where $X$ is a Gaussian Mat\'ern field as described above and $\vep_i$ again is Gaussian measurement noise. 

For the Gaussian and transformed Gaussian model the parameters are estimated using direct optimization of the likelihood given $\mv{y}$ or $\sqrt{\mv{y}}$, as described in \cite{bolin09b}. The model parameters for the GAL and NIG models are estimated using the MCEM procedure described in Section~\ref{sec:estimation}.
The estimates of the parameters for all the models can be seen in Table \ref{tab:parest}.
\begin{table}[t]
\centering
\begin{tabular}{@{}lcccccccc@{}}
\toprule
& \multicolumn{4}{c}{January} & \multicolumn{4}{c}{June}\\
 & tGauss & Gauss & NIG & GAL   & tGauss & Gauss & NIG & GAL \\
\cmidrule(r){2-5} \cmidrule(r){6-9}
$\kappa$ 		& 0.79 	& 0.61 	& 1.63 		& 1.50 	& 0.71 	& 0.23 	& 1.5 	& 1.10\\
$\phi$ 			& 10.13 & 253 	& - 		& -		& 7.44 	& 135 	& - 	& -	\\
$\sigma_{\vep}$ & 1.37 	& 32.5	& 20.3		& 23.4	& 1.36 	& 29.0 	& 25.6 	& 27.2	\\
$\beta_1$		& 7.57 	& 5.14 	& 32		& -12.5	& 8.26 	& 3.44 	& 12.5 	& -206 \\
$\mu$ 			& - 	& - 	& 5.3e3		& 53.5	& -    	& - 	& 429 	& 66.8	\\
$\sigma$ 		& - 	& - 	& 5.3e3		& 157.6	& - 	& - 	& 302	& 28.5 \\
$\nu^2$ 		& - 	& - 	& 4.3e-5	& -		& - 	& - 	& 0.02 	& -	\\
$\tau$ 			& - 	& - 	& -			& 3.6	& - 	& - 	& -		& 5.22	\\
\bottomrule\noalign{\smallskip}
\multicolumn{9}{c}{(a) Without PRISM} \\[0.3cm]
\toprule
& \multicolumn{4}{c}{January} & \multicolumn{4}{c}{June}\\
 & tGauss & Gauss & NIG & GAL   & tGauss & Gauss & NIG & GAL \\
\cmidrule(r){2-5} \cmidrule(r){6-9}
$\kappa$ 		& 0.72 	& 0.79 	& 1.13 		& 1.55  & 1.18 	& 1.26 	& 1.22   	& 1.47 \\
$\phi$ 			& 4.18 	& 107.1 & - 		& - 	& 7.58 	& 147.8 & - 	   	& - 	\\
$\sigma_{\vep}$ & 0.99 	& 21.6  & 17.0 		& 17.9	& 1.29 	& 27.1 	& 16.7	   	& 25.6 	\\
$\beta_1$		& -1.10	& -5.27 & -15.0		& -35.0	& 0.26	& -0.12 & -16.3 	& -86.2 \\
$\beta_2$		& 1.18 	& 1.36  & 1.30		& 1.27	& 0.99 	& 1.10 	& 1.28	   	& 0.84	\\
$\mu$ 			& - 	 & - 	& -19.7		& 16.2 	& - 	& - 	& 245.7 	& 43.5 	\\
$\sigma$ 		& - 	 & - 	& 1868 		& 79.0 	& -	 	& - 	& 1169 		& 69.0 	\\
$\nu^2$ 		& - 	 & - 	& 4.3e-5 	& - 	& - 	& - 	& 1.5e-4 	& - 	\\
$\tau$ 			& - 	 & - 	& - 		& 4.0	& - 	& - 	& - 		& 5.26	\\
\bottomrule\noalign{\smallskip}
\multicolumn{9}{c}{(b) Using PRISM}
\end{tabular}
\caption{Parameter estimates for the different models for the precipitation data from January and June. Note that the Gaussian parameters are for transformed data while NIG and GAL parameters are for raw data and hence should not be compared directly. This is, for example, the reason for the large differences in measurement noise variances.}
\label{tab:parest}
\end{table}
\subsubsection{Models with a PRISM covariate for the mean}
The models described above used no covariates. It is not easy to find good covariates for precipitation modeling; however, \cite{johns2003infilling} used a climate estimate for precipitation obtained using the PRISM method \citep{gibson1997derivation,daly1994statistical} as a covariate when analysing the dataset. The PRISM covariate explains much of the variation in the data, but it should be noted that it is partially based on the same measurements as we are studying. 

The PRISM covariate is included in the models as a covariate for the mean value. Thus, for the untransformed models, $\beta$ is replaced with 
$\beta_1 + \beta_2 B_2(\mv{s})$, and for the transformed Gaussian model, $\beta$ is replaced with $\beta_1 + \beta_2 \sqrt{B_2(\mv{s})}$, where $B_2(\mv{s})$ is the PRISM covariate. The parameters are estimated in the same way as for the models without the PRISM covariate and the parameter estimates for all models with the PRISM covariate can be seen in Table \ref{tab:parest}.

\begin{figure}[t]
\begin{center}
\begin{minipage}[b]{0.4\linewidth}
\centering
January\\
\input{figs/precip_gauss_jan_qq.tex}
\includegraphics[width=\linewidth,height=4.2cm]{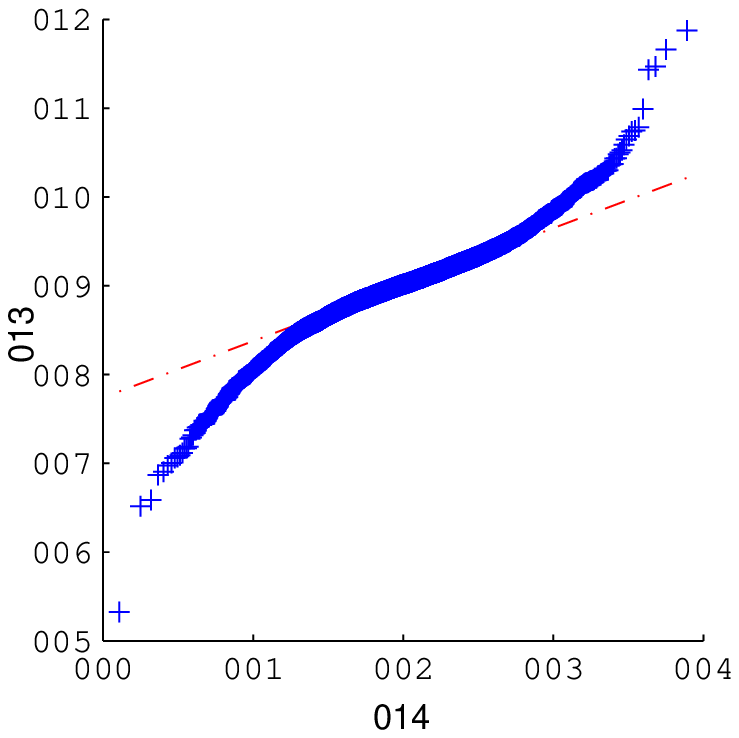}%
\end{minipage}
\begin{minipage}[b]{0.4\linewidth}
\centering
June\\
\input{figs/precip_gauss_jun_qq.tex}
\includegraphics[width=\linewidth,height=4.2cm]{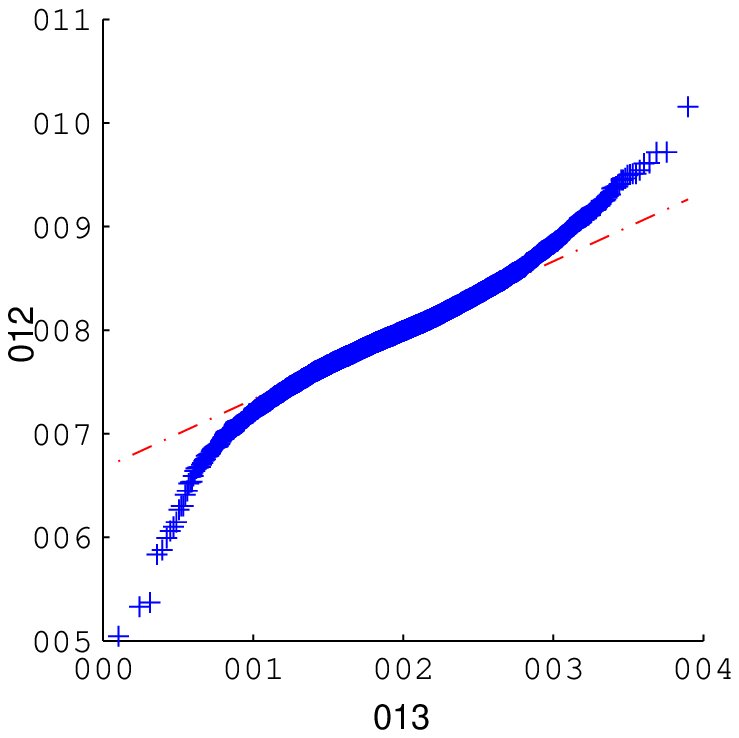}%
\end{minipage}
\end{center}
\vspace{-0.4cm}
\caption{Quantile-quantile plots of standardized residuals for the transformed Gaussian model. A clear deviation from gaussianity can be seen for both months.}
\label{fig:qqgaussian}
\end{figure}

Before going further into the analysis of the different models and the data, it might be of interest to see if there are any reasons for considering anything else than a transformed Gaussian model with PRISM as a covariate. If this model was correct,  $(\sqrt{\mv{y}}-\pE(\mv{X}|\mv{y},\ParFamily))/(\pV(\mv{X}|\mv{y},\ParFamily) + \sigma_{\vep}^2)^{1/2}$ should approximately follow a standard Gaussian distribution if $\mv{X}$ is the latent field evaluated at the measurement locations. Quantile-quantile plots of these standardized prediction residuals are shown in Figure \ref{fig:qqgaussian}. As seen in the figure, the Gaussian fit is far from perfect for both June and January which serves as a motivation for considering the other models.

\subsection{Model selection using cross-validation}
A natural question is which model that fits the data best. Since the main goal of the analysis is to do spatial infilling, we focus the model comparison on the accuracy of the spatial predictions and their corresponding error estimates. To compare the different models ability to do spatial prediction, we use cross-validation. The dataset is divided into ten equally large groups $\mv{y}_1, \ldots, \mv{y}_{10}$, by doing a random permutation of the dataset and then choosing the first tenth of the dataset as $\mv{y}_1$, the second tenth as $\mv{y}_2$, etc. For each $k=1,\ldots,10$, the expectations $\pE(\mv{y}_k|\mv{y}_{(-k)})$ and the variances $\pV(\mv{y}_k|\mv{y}_{(-k)})$ are calculated, which are the spatial predictions and their variances for the locations in group $k$ using all data except the data in that group. By calculating these values for all groups, predictions are performed at all measurement locations, and by subtracting the measurements from these predictions we obtain a complete set of cross-validation residuals $\mv{r}$. By dividing each value in $\mv{r}$ with the predicted kriging variance for that location, we obtain a set of standardized residuals $\mv{r}_s$ which should have variance one if the model is correct. 

In Table \ref{tab:crossval}, the estimated variance of the standardized residuals, $\pV(\mv{r}_s)$, is reported,  together with several other measures of the residuals: The estimated mean, $\pE(\mv{r})$, which should be close to zero, the estimated variance, $\pV(\mv{r})$, which should be as small as possible, the estimated mean of the absolute values, $\pE(|\mv{r}|)$, which should be close to zero, as well as the continuous ranked probability score (CRPS) \citep{matheson1976scoring} and the energy score, $\|es\|$ \citep{gneiting2008assessing}.
To calculate $CRPS$, we define $\hat{Y}_i,\hat{Y}^{(1)}_i,\hat{Y}^{(2)}_i$ as independent random variables with distribution $\pi(y_i|\mv{y}_{-k(i)})$ where $k(i)$ is the group that observation $i$ belongs to, then
\begin{align}\label{eq:CRPS}
CRPS = m^{-1}\sum_{i=1}^m \E[|y_i - \hat{Y}_i|] + E[|\hat{Y}^{(1)}_i - \hat{Y}^{(2)}_i|].
\end{align}
 The CRPS is the most employed scoring role in probabilistic forecasts and the energy score is a multivariate extension of the CRPS. 

\begin{table}[t]
\centering
\begin{tabular}{@{}lcccccccc@{}}
\toprule
& \multicolumn{4}{c}{January} & \multicolumn{4}{c}{June}\\
 & tGauss & Gauss & NIG & GAL   & tGauss & Gauss & NIG & GAL \\
\cmidrule(r){2-5} \cmidrule(r){6-9}
$\pV(\mv{r}_s)$	& $1.42$ 	& $1.46$ & $1.7$ 	  & $\mv{1.38}$  & $0.78$ & $3.19$  &  $1.10$ &  $\mv{1.06}$ \\
$\pE(\mv{r})$	& $3.6$	& $\mv{0.13}$ & $-1.00$ 	  & $-0.50$  & $2.12$ & $\mv{-0.05}$ 		& $-0.06$ & $0.18$		\\
$\pV(\mv{r})$	& $1423$	& $\mv{1415}$ & $2494$   & $1436$  & $1048$ & $1046$  & $1087$  & $\mv{1036}$ \\
$\pE(|\mv{r}|)$ & $21.5$  & $20.5$ & $22.4$   & $\mv{20.3}$  & $22.9$ & $22.4$  & $22.7$  & $\mv{22.4}$  \\ 
$||es||$        & $2221$  & $2190$ &  $2496$  &  $\mv{2175}$ & $1925$ & $1951.0$  & $1922$  &  $\mv{1874}$ \\ 
$CRPS$          & $\mv{15.6}$	& $17.8$ & $18.0$ & $16.8$  & $\mv{16.30}$ & $17.48$  & $17$    &  $16.9$ \\ 
\bottomrule\noalign{\smallskip}
\multicolumn{9}{c}{(a) Without PRISM} \\[0.3cm]
\toprule
& \multicolumn{4}{c}{January} & \multicolumn{4}{c}{June}\\
 & tGauss & Gauss & NIG & GAL   & tGauss & Gauss & NIG & GAL \\
\cmidrule(r){2-5} \cmidrule(r){6-9}
$\pV(\mv{r}_s)$		  	& $0.72$ 	& $2.02$		& $1.33$ 	& $\mv{1.16}$ 	& $0.88$  		& $2.90$ 		& $2.3$ 	& $\mv{1.02}$	\\
$\pE(\mv{r})$		  	& $0.70$	& $0.17$		& $-0.28$	& $\mv{-0.07}$ 	& $0.95$		& $\mv{0.01}$ 	& $0.57$	& $\mv{-0.01}$	\\
$\pV(\mv{r})$		  	& $\mv{520}$ 	& $575$			& $873$ 	& $569$  	& $962$   	& $967$  		& $1317$	& $\mv{961}$	\\
$ \pE(|\mv{r}|)$      	& $\mv{13.7}$  	& $1.43$ 		& $14.9$ 	& $14.1$    &  $21.3$ 	& $21.2$ 		& $22.7$ & $\mv{21.2}$  \\ 
$||es||$              	& $\mv{1333}$  	& $1385$ 		& $1717$	& $1360$    & $1834$ 		& $1833$ 		& $2201$  &  $\mv{1809}$ \\ 
$CRPS$        			& $\mv{10.1}$	&$11.8$ 		& $11.8$ 	& $11.5$    & $\mv{15.4}$ 	& $16.4$ 		& $17.3$     &  $16.3$ \\ 
\bottomrule\noalign{\smallskip}
\multicolumn{9}{c}{(b) With PRISM}
\end{tabular}
\caption{Crossvalidation results for the different models. Here, $\mv{r}$ denotes the actual model residuals and $\mv{r}_s$ denotes the same residuals standardized by the estimated kriging variances. $||es||$ denotes the energy norm of $\mv{r}$ and $CRPS$ denotes the continuous ranked probability score of $\mv{r}$. The best value for each month is marked with bold script.}
\label{tab:crossval}
\end{table}

There are several things to note in the tables. First of all, the mean and variance are similar for all models, except for NIG model. The reason for the large variance for the NIG model could be that the estimation puts too much emphasis on allowing for big jumps in the variance process to account for outliers in the data, and as a result, the tails of the distribution fits the data well but the fit in general is poor. Also, a peculiar effect for the NIG model is that the addition of the PRISM covariate actually worsens cross-validation results for the June data, whereas it improves the performance for all other models. After further analysis of the NIG results, we found that the addition of the PRISM covariate reduces the size of the residuals when fitting the model to the whole dataset but increases cross-validation residuals, which indicates overfitting and the NIG model is therefore likeliy not a suitable model for this dataset. 

The Gaussian model severely underestimates the variance of the Kriging estimator, which can be seen in the variance of the standardized residuals.  Overall, the transformed Gaussian model and the GAL model seems to preform the best, both with and without the PRISM covariate, and we therefore choose two study the difference between these two models in more detail.

\subsection{A comparison between the GAL and transformed Gaussian models}
\begin{figure}[t]
\begin{center}
\centering 
GAL\\
\begin{minipage}[b]{0.4\linewidth}
\centering
January 1997\\
\includegraphics[width=\linewidth,bb= -0 -0 451 202,clip=]{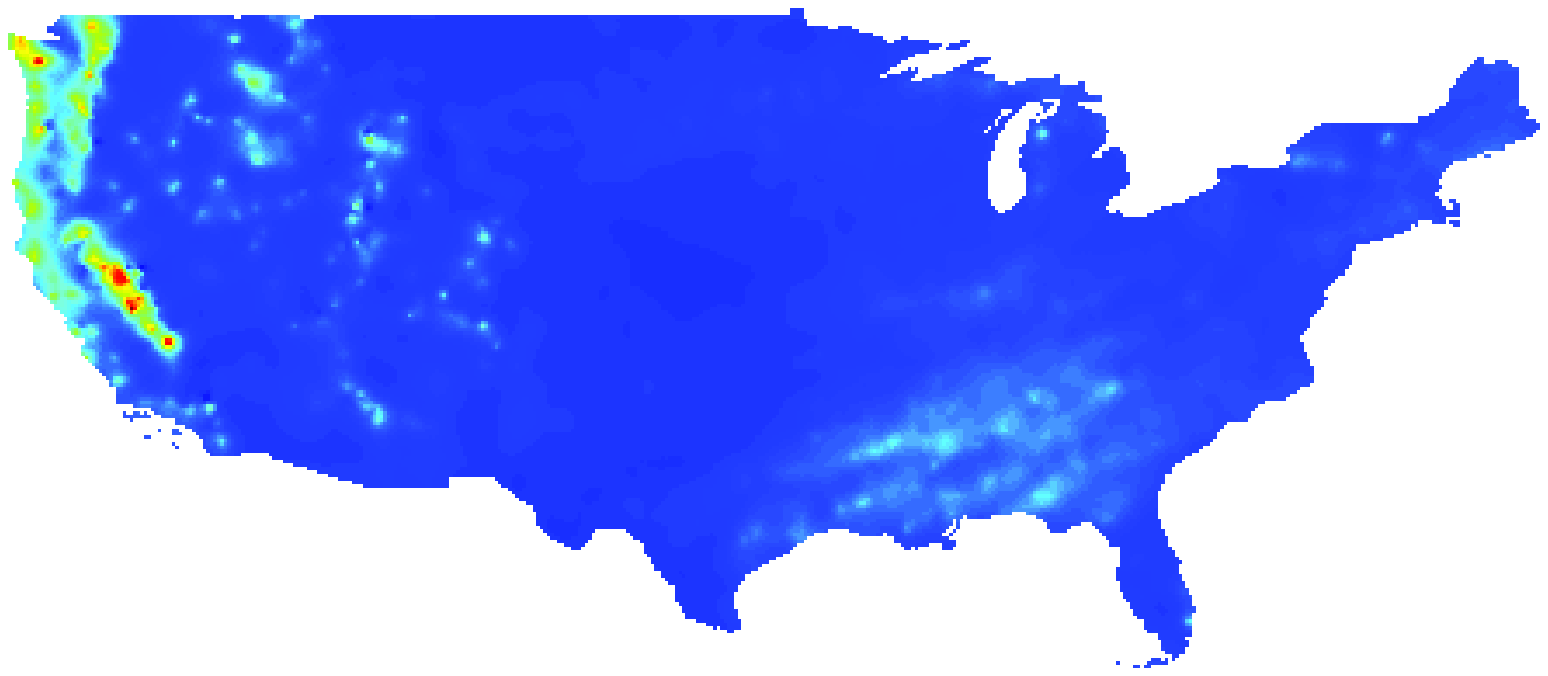}%
\end{minipage}
\begin{minipage}[b]{0.07\linewidth}
\centering
\quad\\
\input{figs/precip_GAL_krig_jan_noprism_cb.tex}
\includegraphics[height=25mm,bb=55 381 100  523,clip=]{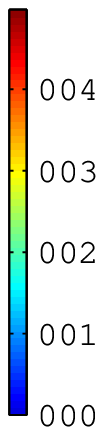}%
\end{minipage}
\begin{minipage}[b]{0.4\linewidth}
\centering
June 1997\\
\includegraphics[width=\linewidth,bb=-0 -0 451 202,clip=]{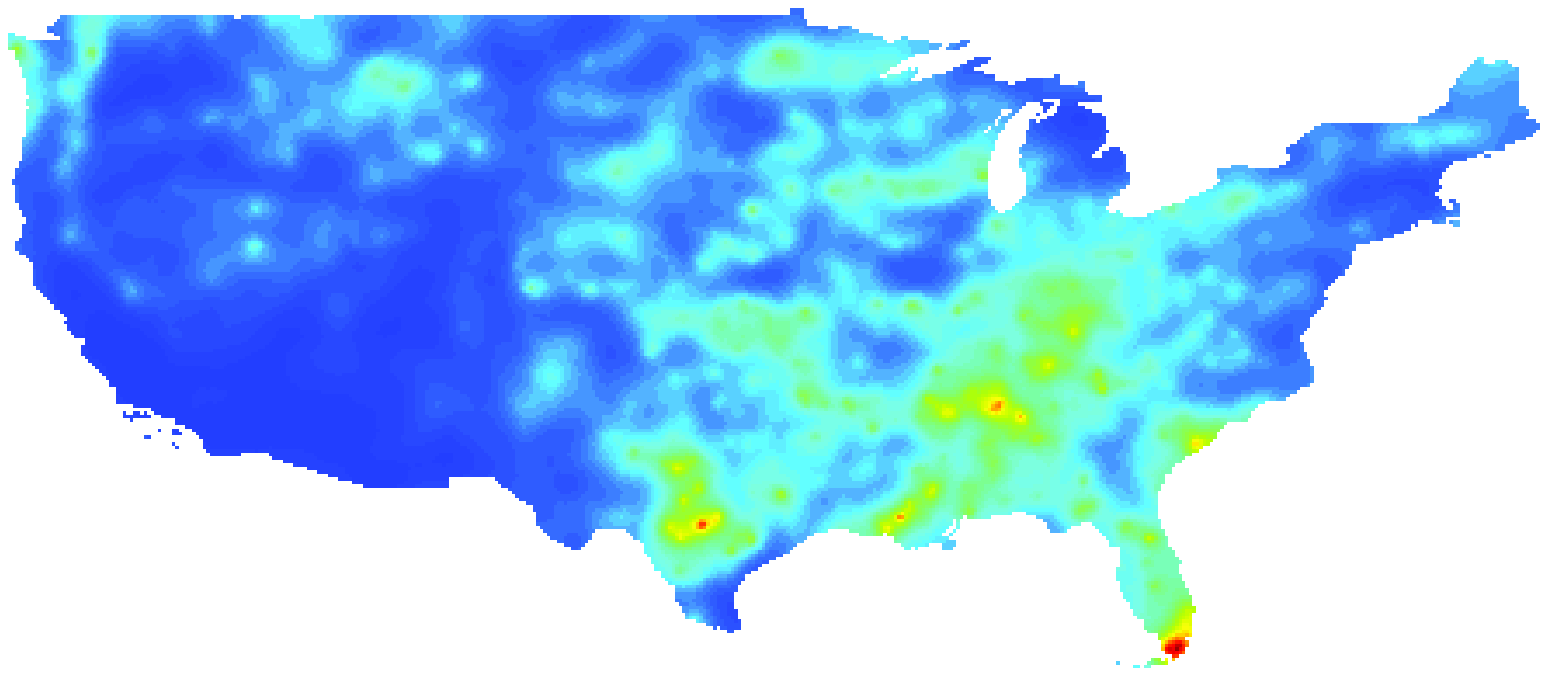}%
\end{minipage}
\begin{minipage}[b]{0.07\linewidth}
\centering
\quad\\
\input{figs/precip_GAL_krig_jun_noprism_cb.tex}
\includegraphics[height=25mm,bb=55 381 100  523,clip=]{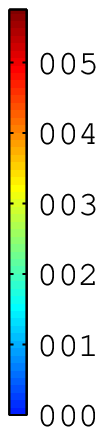}%
\end{minipage}
\centering 
Transformed Gaussian \\
\begin{minipage}[b]{0.4\linewidth}
\centering
\includegraphics[width=\linewidth,bb= -0 -0 451 202,clip=]{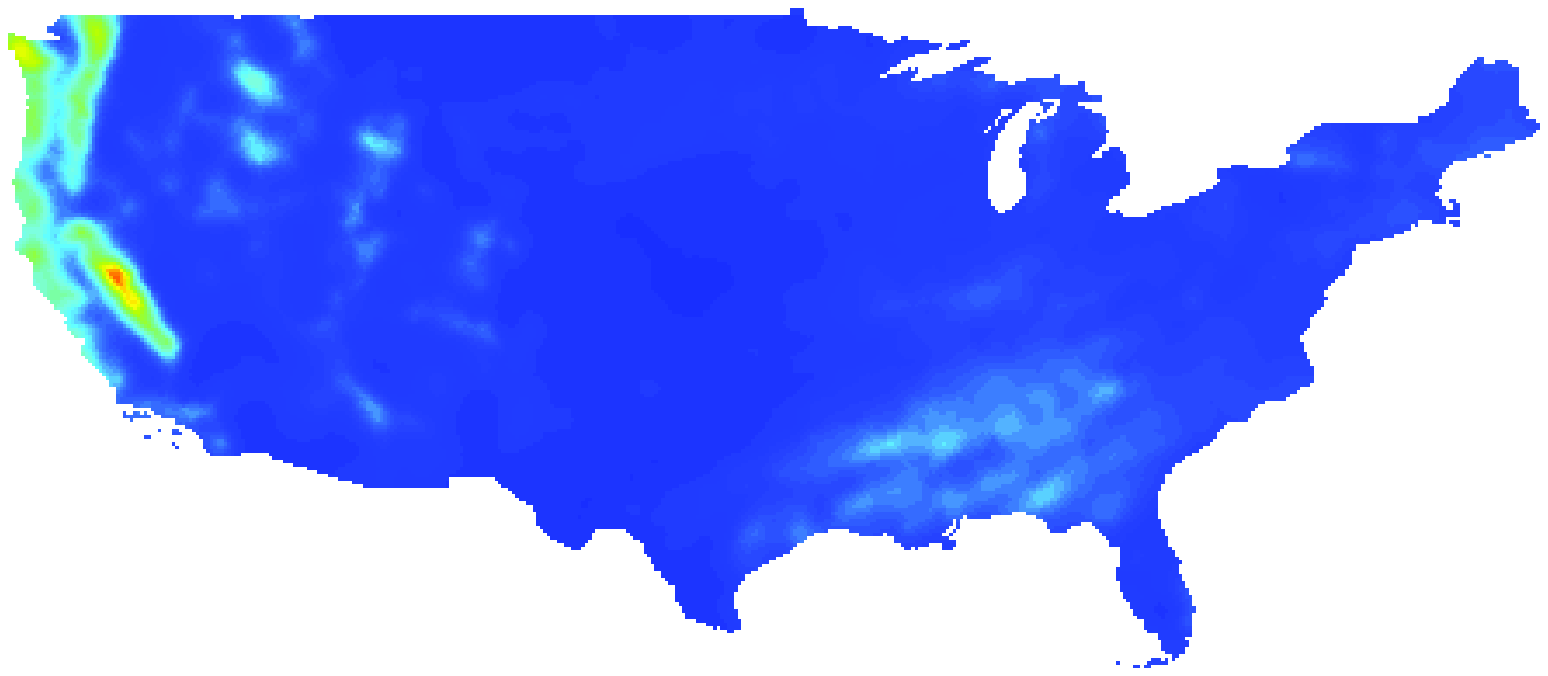}%
\end{minipage}
\begin{minipage}[b]{0.07\linewidth}
\centering
\quad\\
\input{figs/precip_tgauss_krig_jan_noprism_cb.tex}
\includegraphics[height=25mm,bb=    55  2836    100  2978,clip=]{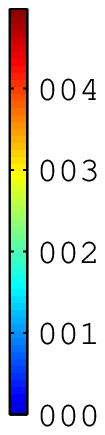}%
\end{minipage}
\begin{minipage}[b]{0.4\linewidth}
\centering
\includegraphics[width=\linewidth,bb= -0 -0 451 202,clip=]{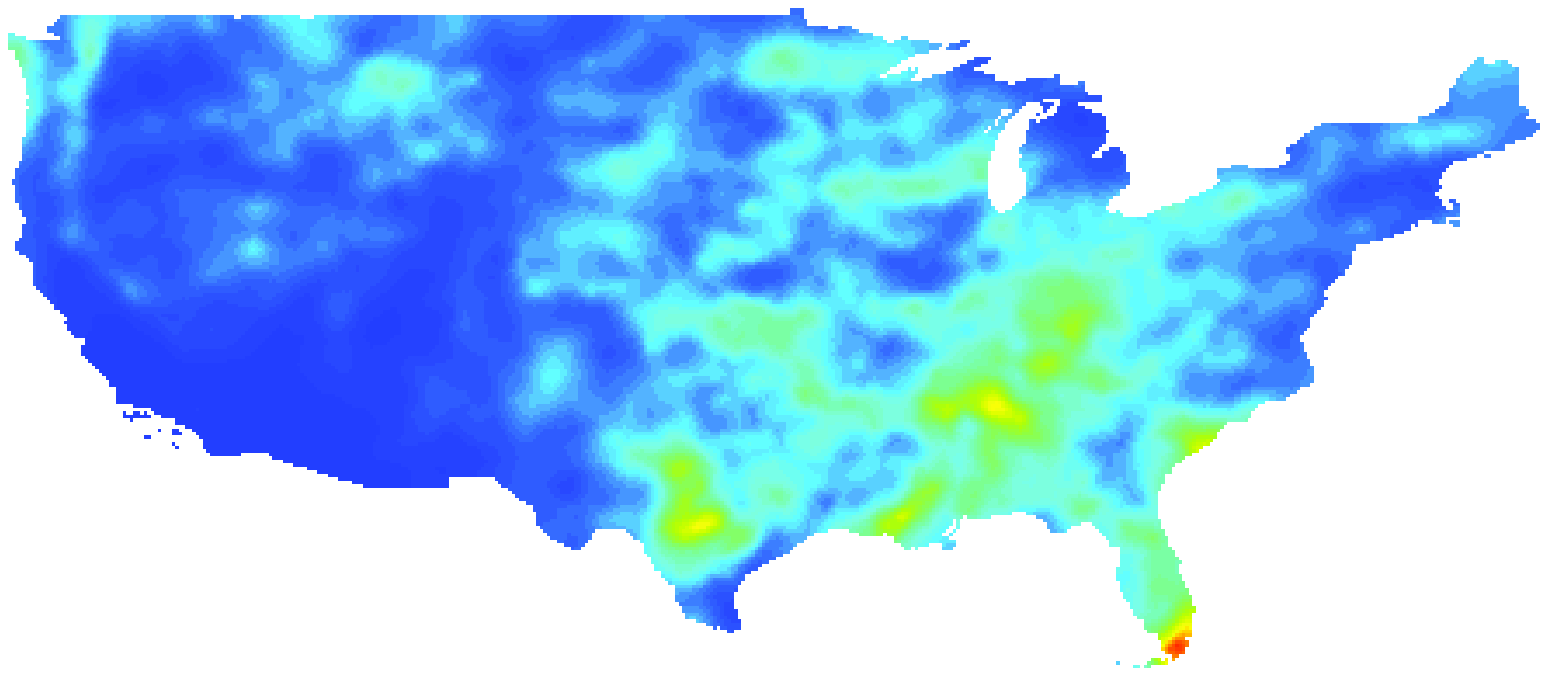}%
\end{minipage}
\begin{minipage}[b]{0.07\linewidth}
\centering
\quad\\
\input{figs/precip_tgauss_krig_jun_noprism_cb.tex}
\includegraphics[height=25mm,bb=55   401    95   523,clip=]{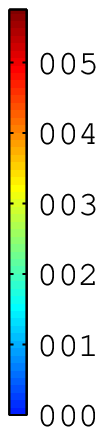}%
\end{minipage}
\centering 
Difference \\
\begin{minipage}[b]{0.4\linewidth}
\centering
\includegraphics[width=\linewidth,bb= -0 -0 451 202,clip=]{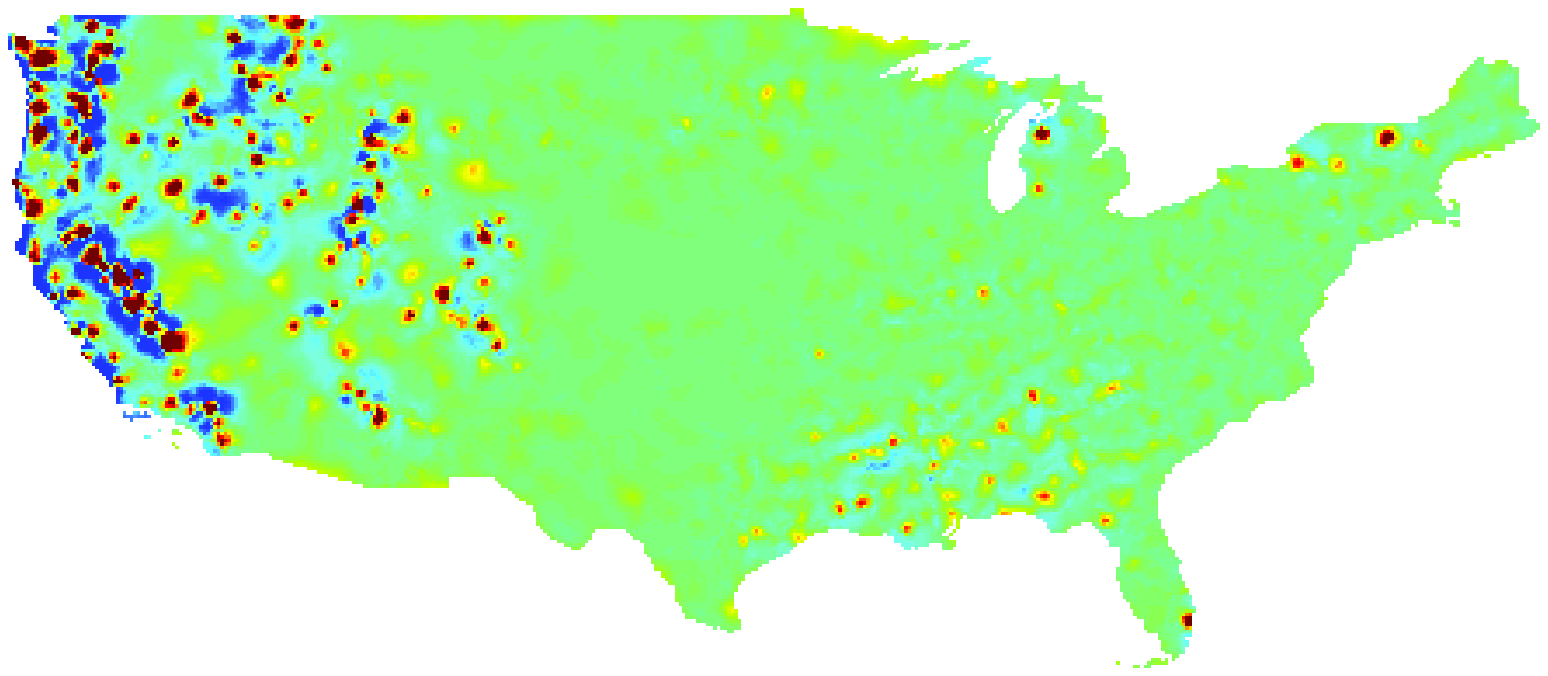}%
\end{minipage}
\begin{minipage}[b]{0.07\linewidth}
\centering
\quad\\
\input{figs/precip_GAL_tguass_krig_jan_noprism_cb.tex}
\includegraphics[height=25mm,bb=    55  381    100  523,clip=]{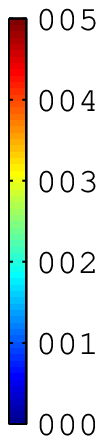}%
\end{minipage}
\begin{minipage}[b]{0.4\linewidth}
\centering
\includegraphics[width=\linewidth,bb= -0 -0 451 202,clip=]{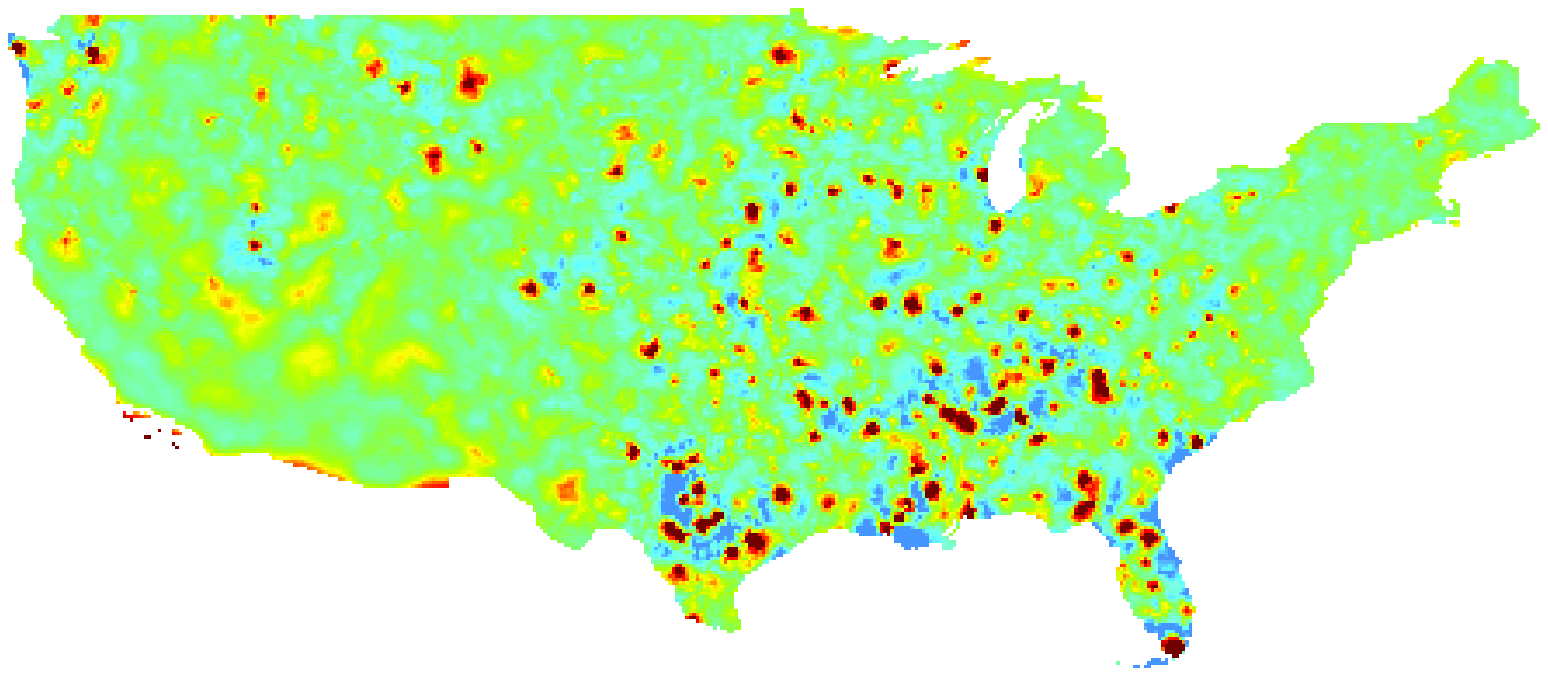}%
\end{minipage}
\begin{minipage}[b]{0.07\linewidth}
\centering
\quad\\
\input{figs/precip_GAL_tguass_krig_jun_noprism_cb.tex}
\includegraphics[height=25mm,bb=    55  381    100  523,clip=]{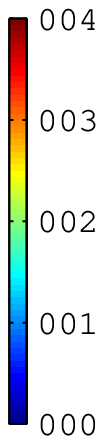}%
\end{minipage}
\end{center}
\vspace{-0.4cm}

\caption{The posterior mean of the fields for January and June 1997 using the GAL model (top), the transformed Gaussian model (mid), and the difference between the two (bottom).}
\label{fig:kriging}
\end{figure}

\begin{figure}[t]
\begin{center}
\centering 
GAL \\
\begin{minipage}[b]{0.4\linewidth}
\centering
January 1997\\
\includegraphics[width=\linewidth,bb= -0 -0 451 202,clip=]{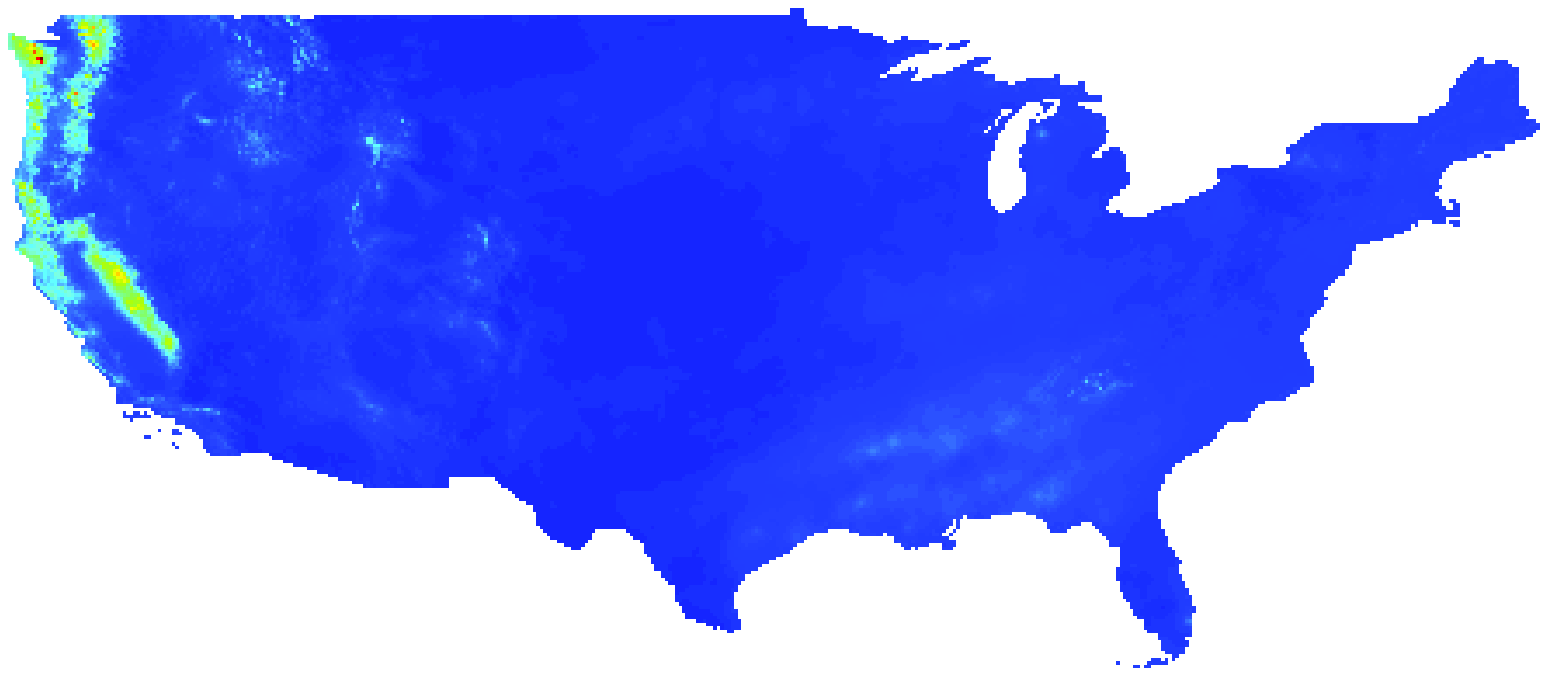}%
\end{minipage}
\begin{minipage}[b]{0.07\linewidth}
\centering
\quad\\
\input{figs/precip_GAL_krig_jan_prism_cb.tex}
\includegraphics[height=25mm,bb=    55  381    100  523,clip=]{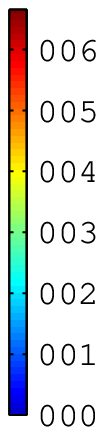}%
\end{minipage}
\begin{minipage}[b]{0.4\linewidth}
\centering
June 1997\\
\includegraphics[width=\linewidth,bb= -0 -0 451 202,clip=]{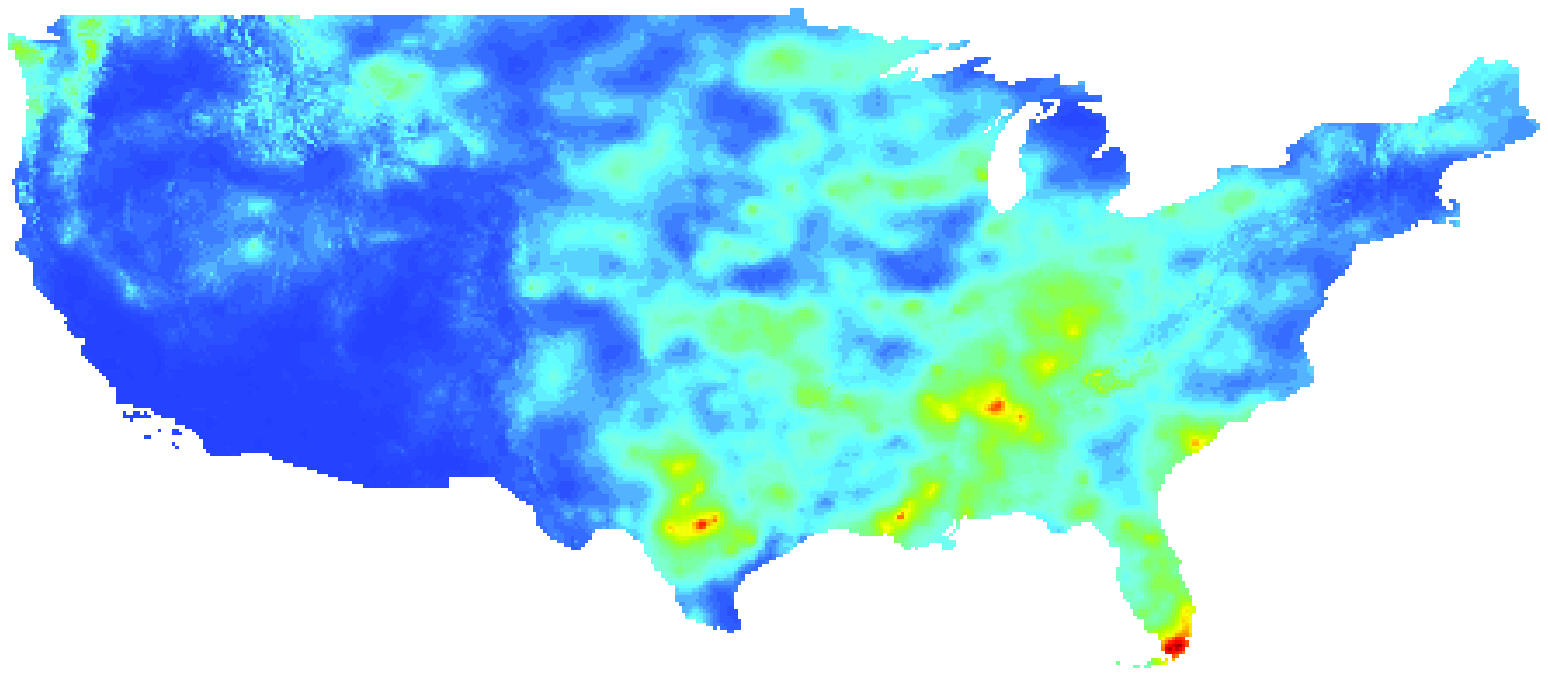}%
\end{minipage}
\begin{minipage}[b]{0.07\linewidth}
\centering
\quad\\
\input{figs/precip_GAL_krig_jun_prism_cb.tex}
\includegraphics[height=25mm,bb=    55  381    100  523,clip=]{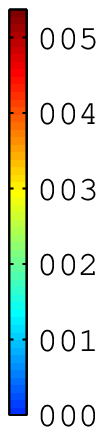}%
\end{minipage}
\centering 
Transformed Gaussian \\
\begin{minipage}[b]{0.4\linewidth}
\centering
\includegraphics[width=\linewidth,bb=-0 -0 451 202,clip=]{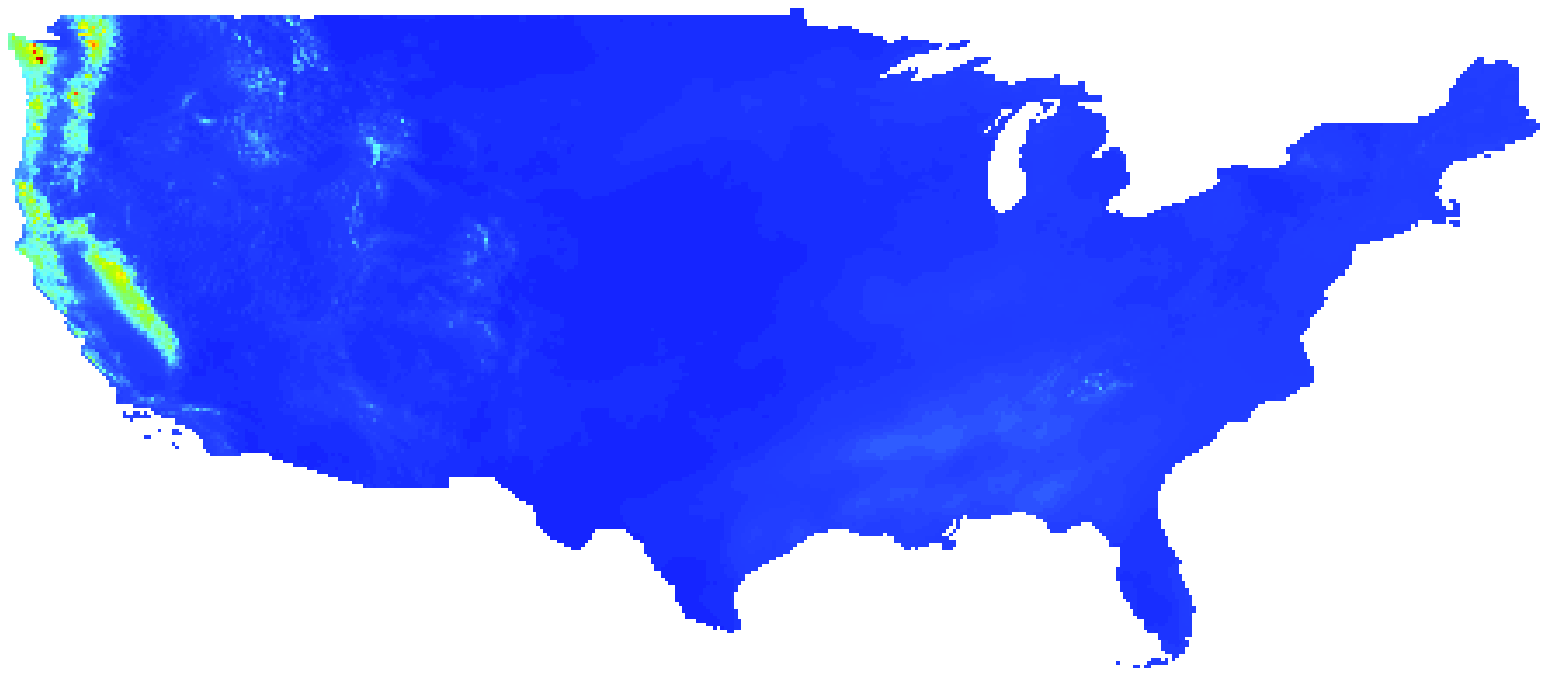}%
\end{minipage}
\begin{minipage}[b]{0.07\linewidth}
\centering
\quad\\
\input{figs/precip_tgauss_krig_jan_prism_cb.tex}
\includegraphics[height=25mm,bb=    55   401    95   523,clip=]{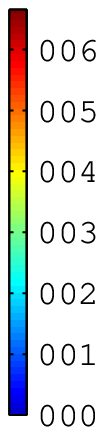}%
\end{minipage}
\begin{minipage}[b]{0.4\linewidth}
\centering
\includegraphics[width=\linewidth,bb= -0 -0 451 202,clip=]{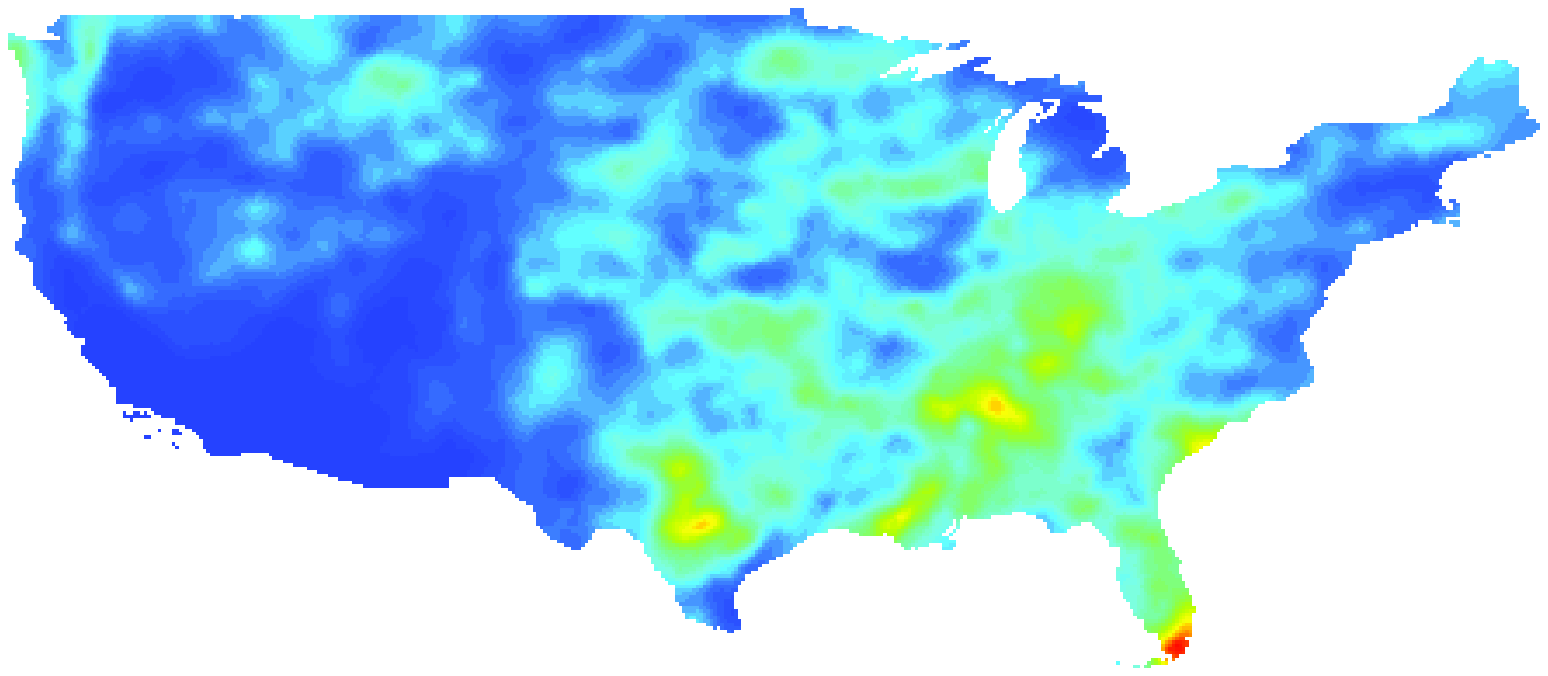}%
\end{minipage}
\begin{minipage}[b]{0.07\linewidth}
\centering
\quad\\
\input{figs/precip_tgauss_krig_jun_prism_cb.tex}
\includegraphics[height=25mm,bb=    55   401    95   523,clip=]{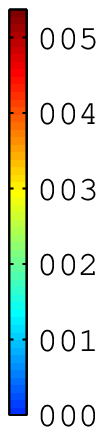}%
\end{minipage}
\quad\\[0.1cm]
\centering 
Difference \\
\begin{minipage}[b]{0.4\linewidth}
\centering
\includegraphics[width=\linewidth,bb= -0 -0 451 202,clip=]{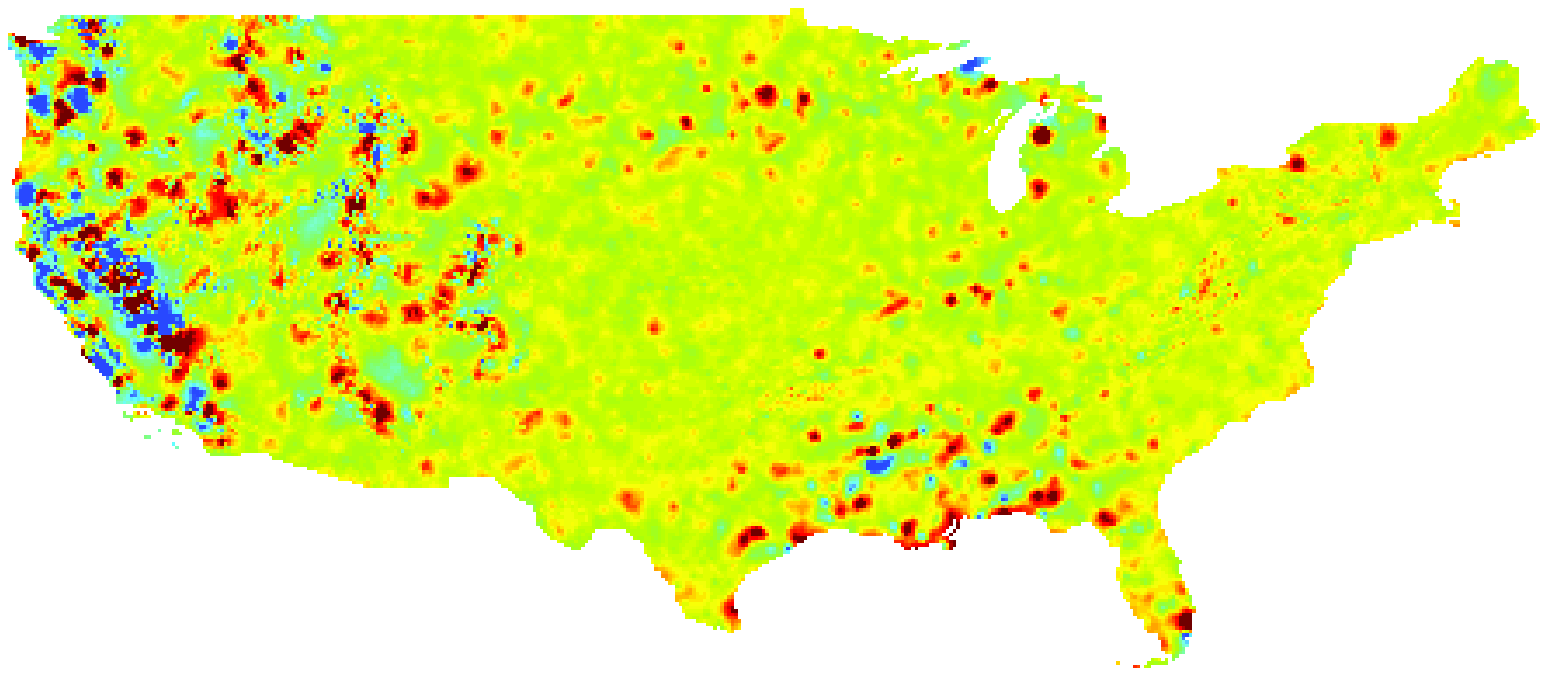}%
\end{minipage}
\begin{minipage}[b]{0.07\linewidth}
\centering
\quad\\
\input{figs/precip_GAL_tguass_krig_jan_prism_cb.tex}
\includegraphics[height=25mm,bb=     55   401    95   523,clip=]{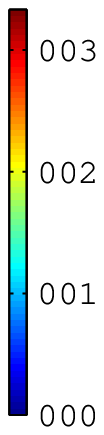}%
\end{minipage}
\begin{minipage}[b]{0.4\linewidth}
\centering
\includegraphics[width=\linewidth,bb= -0 -0 451 202,clip=]{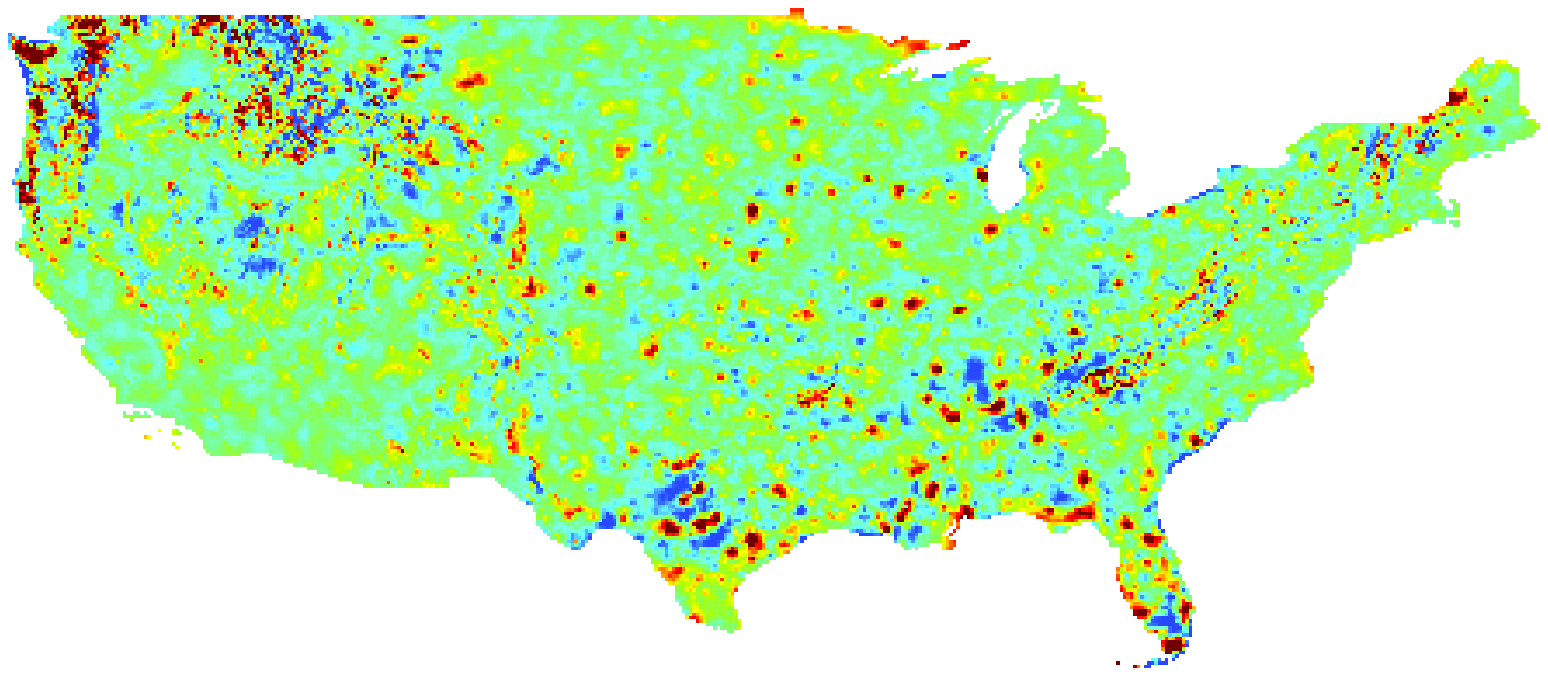}%
\end{minipage}
\begin{minipage}[b]{0.07\linewidth}
\centering
\quad\\
\input{figs/precip_GAL_tguass_krig_jun_prism_cb.tex}
\includegraphics[height=25mm,bb=    55  381    100  523,clip=]{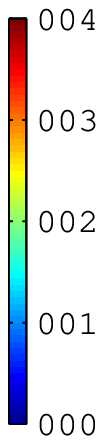}%
\end{minipage}
\end{center}
\vspace{-0.4cm}
\caption{The posterior mean of the fields using the PRISM covariate for January and June 1997 using the GAL model (top), the transformed Gaussian model (mid), and the difference between the two (bottom).}
\label{fig:kriging_prism}
\end{figure}

The posterior mean for the GAL model and the transformed Gaussian model as well as the difference between the two can be seen in Figure \ref{fig:kriging} for the models without the PRISM covariate and in Figure \ref{fig:kriging_prism} for the models with the PRISM covariate. As seen in the figures, the posterior means are not that different between the models. The GAL model has more distinct peaks around the extremes whereas estimate using the transformed Gaussian model is smoother, but the overall pictures are very similar. 

\begin{figure}[t]
\begin{center}
\centering
GAL \\
\begin{minipage}[b]{0.4\linewidth}
\centering
January 1997\\
\includegraphics[width=\linewidth,bb= -0 -0 452 198,clip=]{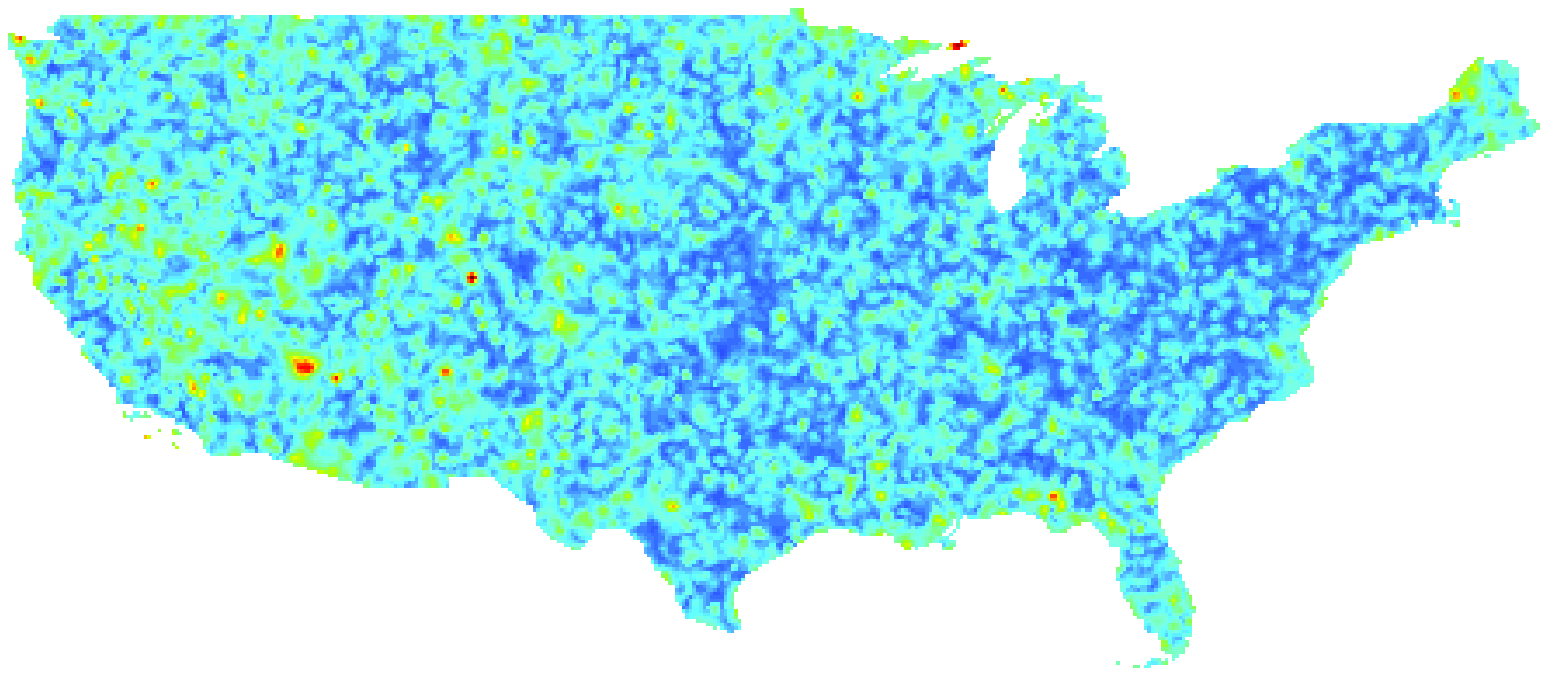}%
\end{minipage}
\begin{minipage}[b]{0.07\linewidth}
\centering
\quad\\
\input{figs/precip_GAL_var_jan_prism_cb.tex}
\includegraphics[height=25mm,bb=    55   401    95   523,clip=]{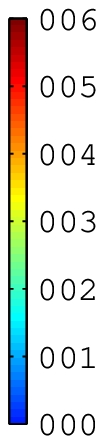}%
\end{minipage}
\begin{minipage}[b]{0.4\linewidth}
\centering
June 1997\\
\includegraphics[width=\linewidth,bb= -0 -0 452 198,clip=]{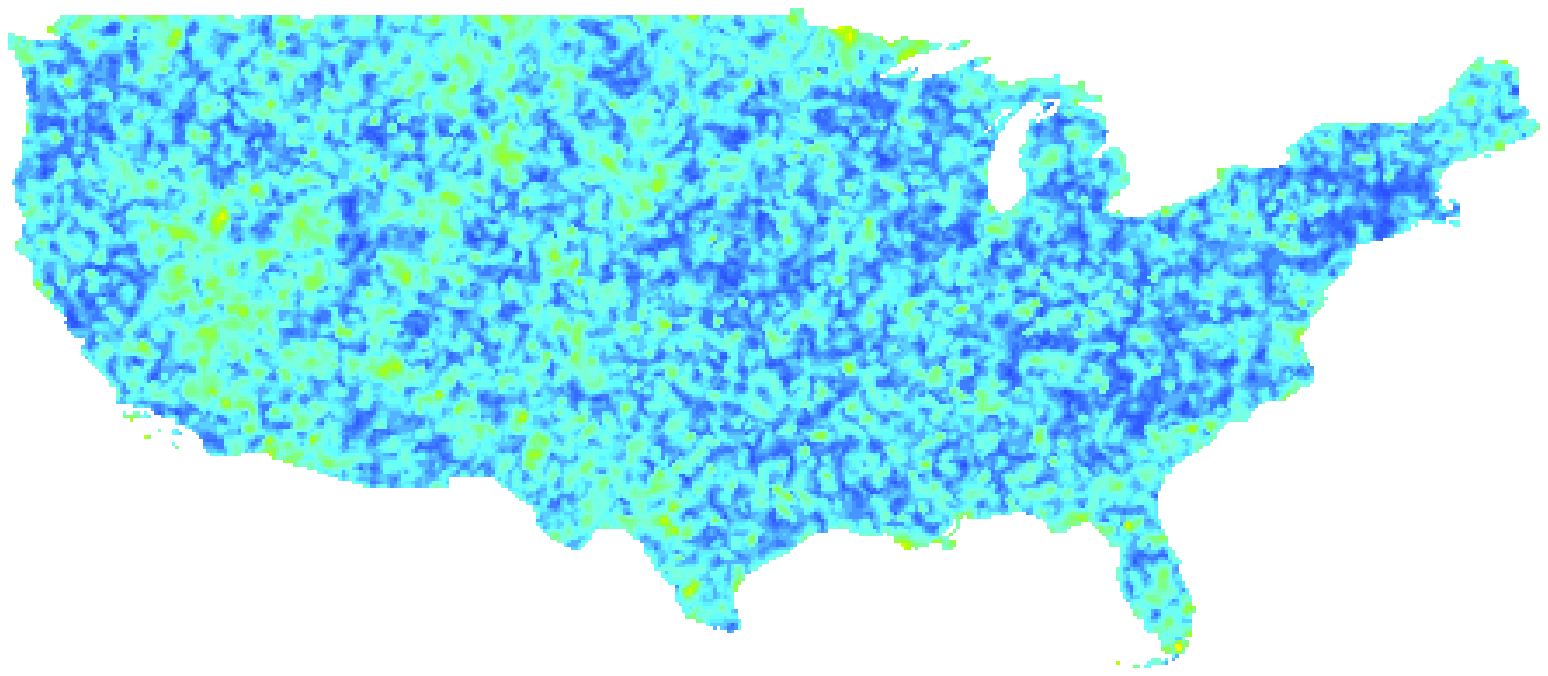}%
\end{minipage}
\begin{minipage}[b]{0.07\linewidth}
\centering
\quad\\
\input{figs/precip_GAL_var_jun_prism_cb.tex}
\includegraphics[height=25mm,bb=   55   401    95   523,clip=]{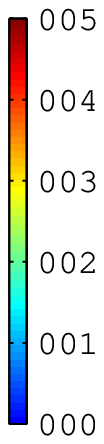}%
\end{minipage}
\quad\\[0.1cm]
\centering
Transformed Gaussian\\
\begin{minipage}[b]{0.4\linewidth}
\centering
\includegraphics[width=\linewidth,bb=-0 -0 452 198,clip=]{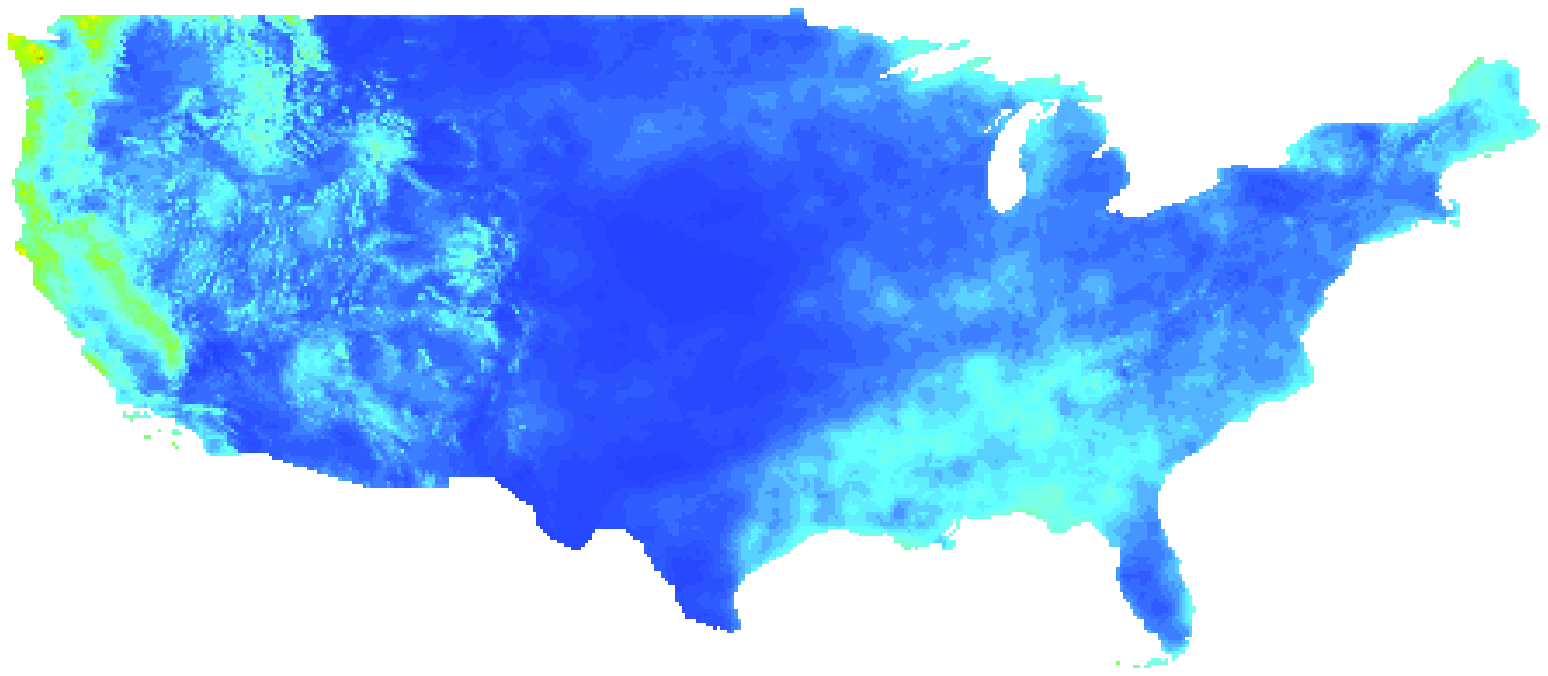}%
\end{minipage}
\begin{minipage}[b]{0.07\linewidth}
\centering
\quad\\
\input{figs/precip_tgauss_var_jan_prism_cb.tex}
\includegraphics[height=25mm,bb= 55  381  95  523,clip=]{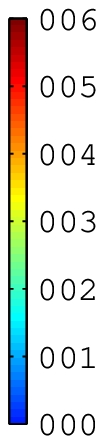}%
\end{minipage}
\begin{minipage}[b]{0.4\linewidth}
\centering
\includegraphics[width=\linewidth,bb= -0 -0 452 198,clip=]{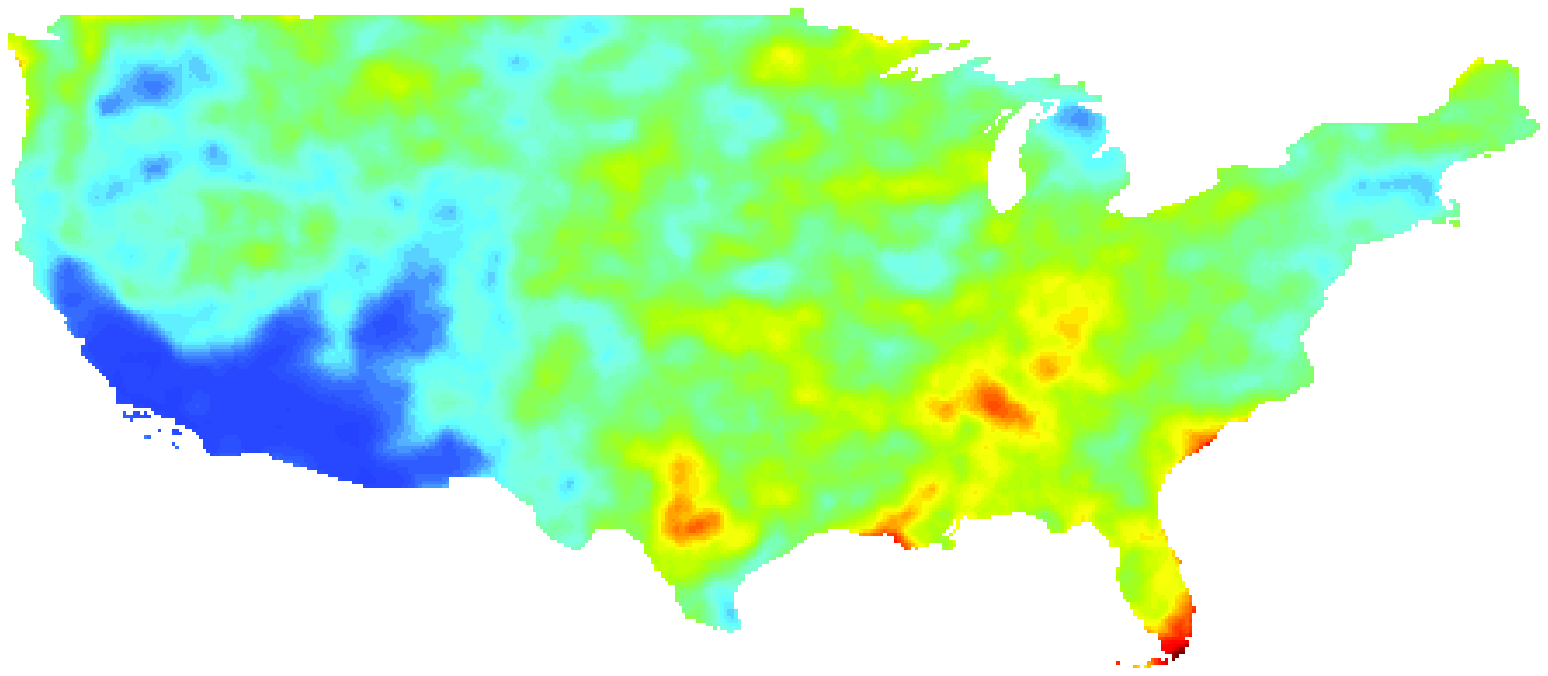}%
\end{minipage}
\begin{minipage}[b]{0.07\linewidth}
\centering
\quad\\
\input{figs/precip_tgauss_var_jun_prism_cb.tex}
\includegraphics[height=25mm,bb= 55  381  95  523,clip=]{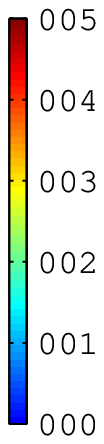}%
\end{minipage}
\end{center}
\vspace{-0.4cm}
\caption{The posterior standard deviations for the GAL model(top) and the transformed Gaussian model  (bottom), for January (left) and June (right) 1997.}
\label{fig:var_tgauss}
\end{figure}
The posterior standard deviations for the transformed Gaussian and the GAL models with the PRISM covariate can be seen in Figure \ref{fig:var_tgauss}. While there was no large difference in the posterior means between the models, the estimates of the posterior variances are completely different between the models. The reason for this can easily be seen by calculating the variance of $X^2$ in the transformed Gaussian model, which is the precipitation in the original scale. Standard calculations give that
\begin{equation*}
\pV[X(\mv{s})^2|\mv{y}] = 2\pV[X(\mv{s})|\mv{y}]^2  + 4 \pE[X(\mv{s})|\mv{y}]^2 \pV[X(\mv{s})|\mv{y}].
\end{equation*}
Hence, the actual values of $\bf y$ affects the kriging variance for the transformed Gaussian model, through the term $\pE[X(\mv{s})|\mv{y}]$, whereas only the distribution of the measurement locations affects the variance for the GAL model. The model's different variance structures are reflected in their respective $CPRS$ in Table \ref{tab:crossval}. There is little precipitation and small variation in the amount of precipitation over large areas in the US. This is effect is modeled best with the transformed Gaussian model, and the second term in \eqref{eq:CRPS} is therefore smaller for the transformed Gaussian model than for the GAL model.  

The reason for the poor values of the standardized variances of the transformed Gaussian model is that the linear interpolation method used in the SPDE method induces a large bias in the variance if a non-linear transformation is used. This can be avoided by using a basis induced by a triangulation with nodes at each observation and prediction location, since this avoids using linear interpolation. Rerunning the cross-validation with nodes at each observation location results in standardized variances close to one for all transformed Gaussian models. However, the goal is to produce high resolution maps of precipitation and adding nodes at each prediction location in these maps results in models that are not computationally feasible to use for either estimation or prediction. Thus, we cannot use the ideal grid for the transformed Gaussian model in practice.

\begin{figure}[t]
\begin{center}
\centering
GAL \\
\begin{minipage}[b]{0.4\linewidth}
\centering
January 1997\\
\includegraphics[width=\linewidth,bb= 0   0   451   202,clip=]{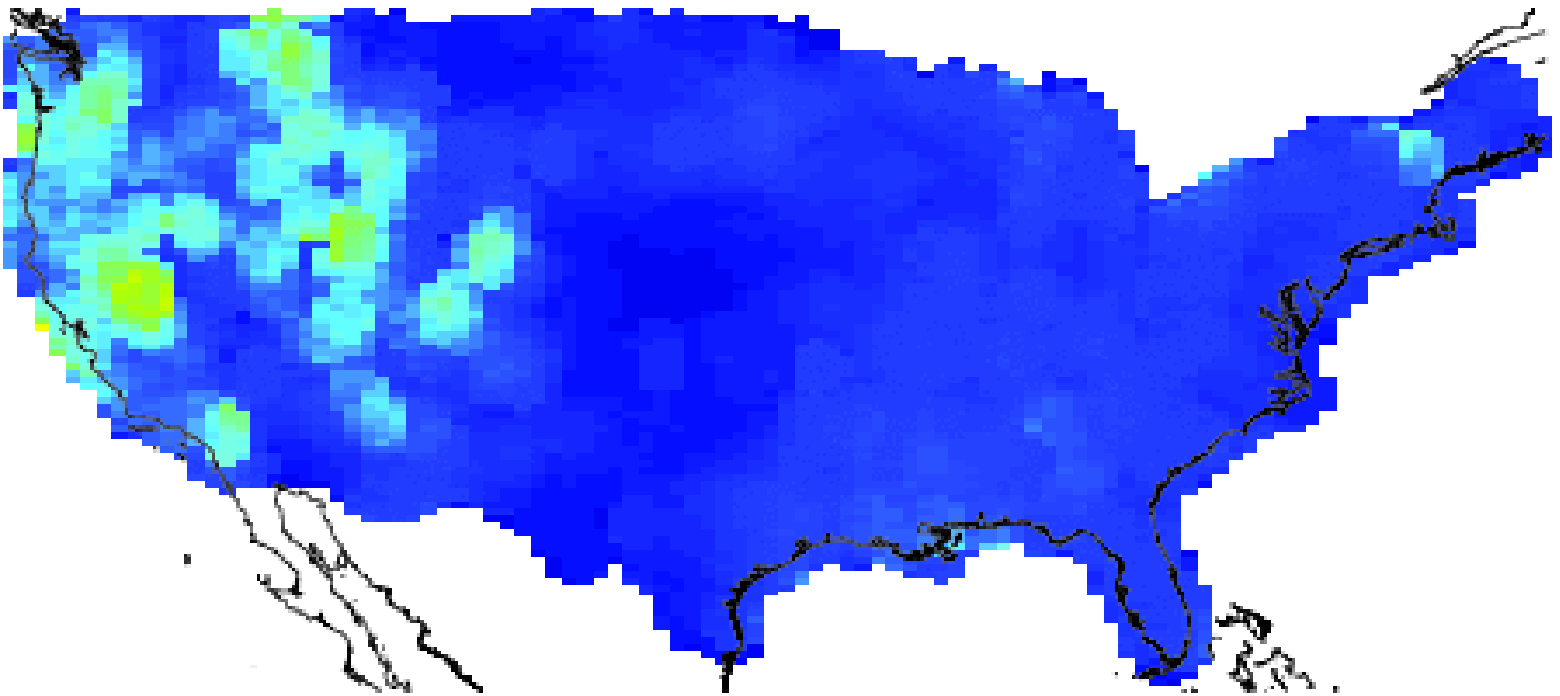}%
\end{minipage}
\begin{minipage}[b]{0.07\linewidth}
\centering
\quad\\
\input{figs/precip_jan_res_gal_sd_cb.tex}
\includegraphics[height=25mm,bb=55  636   102  778,clip=]{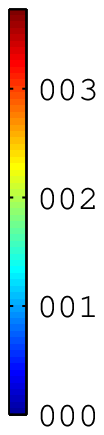}%
\end{minipage}
\begin{minipage}[b]{0.4\linewidth}
\centering
June 1997\\
\includegraphics[width=\linewidth,bb= 0   0   451   202,clip=]{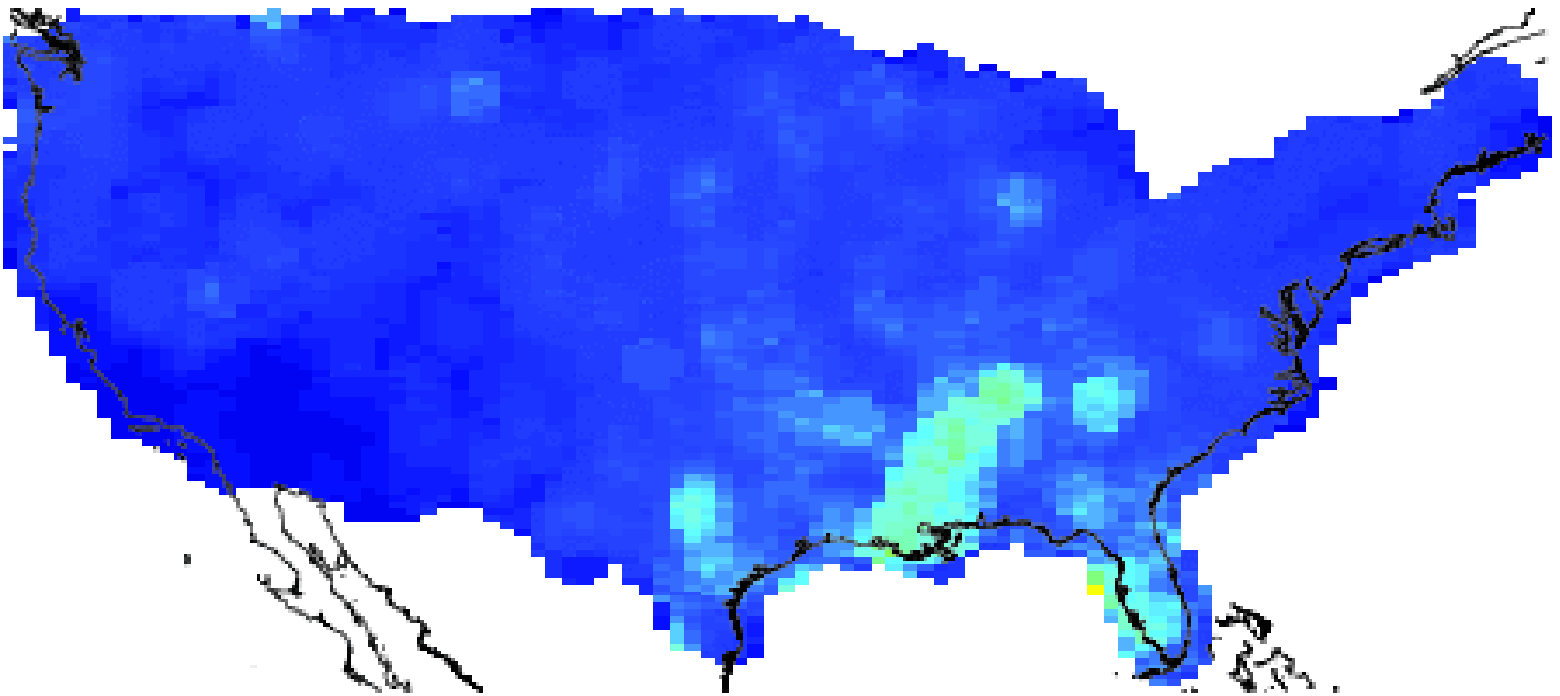}%
\end{minipage}
\begin{minipage}[b]{0.07\linewidth}
\centering
\quad\\
\input{figs/precip_jun_res_gal_sd_cb.tex}
\includegraphics[height=25mm,bb= 55  2836   102  2978,clip=]{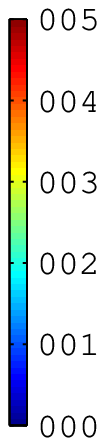}%
\end{minipage}
\quad\\[0.1cm]
\centering
Transformed Gaussian\\
\begin{minipage}[b]{0.4\linewidth}
\centering
\includegraphics[width=\linewidth,bb=  0   0   451   202,clip=]{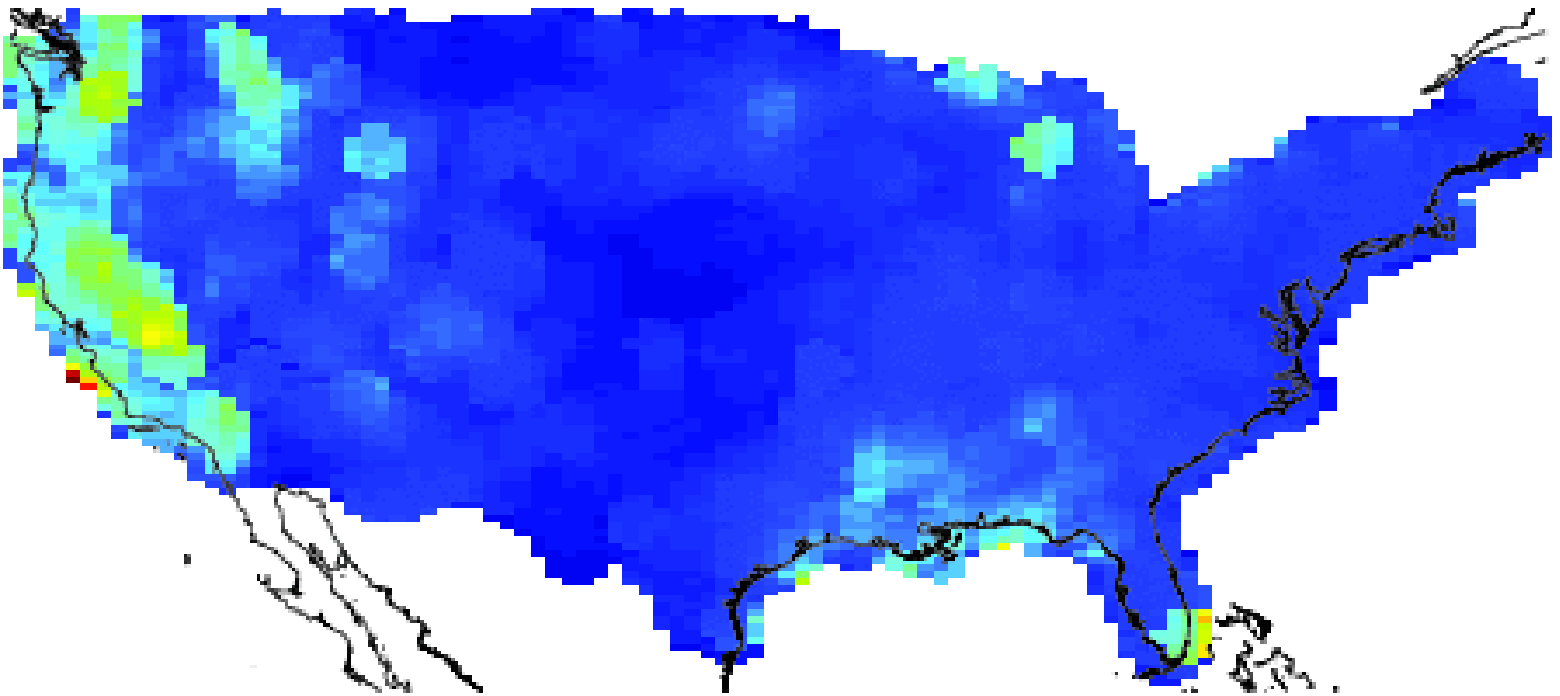}%
\end{minipage}
\begin{minipage}[b]{0.07\linewidth}
\centering
\quad\\
\input{figs/precip_jan_res_tgauss_sd_cb.tex}
\includegraphics[height=25mm,bb= 55  636   102  778,clip=]{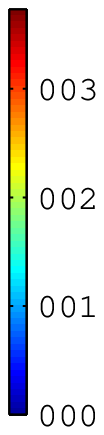}%
\end{minipage}
\begin{minipage}[b]{0.4\linewidth}
\centering
\includegraphics[width=\linewidth,bb=  0   0   451   202,clip=]{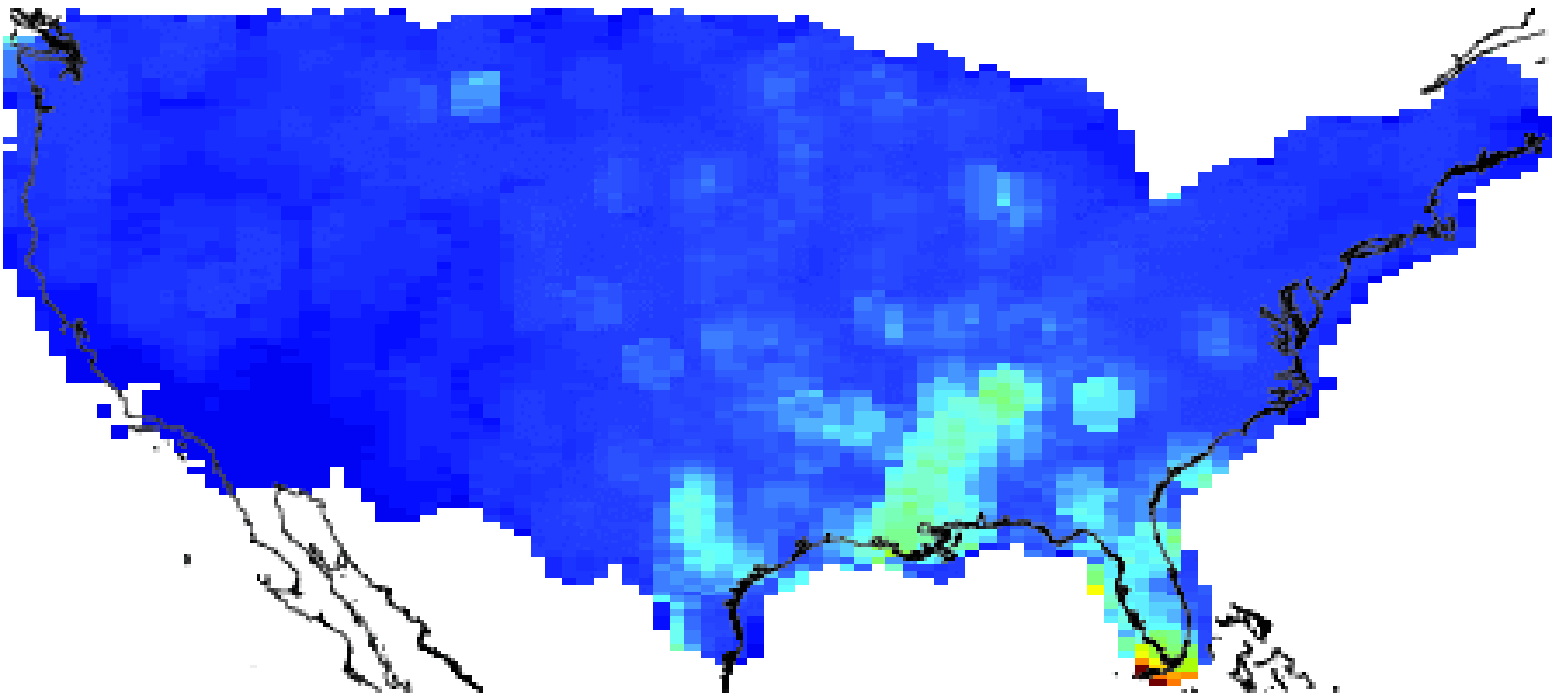}%
\end{minipage}
\begin{minipage}[b]{0.07\linewidth}
\centering
\quad\\
\input{figs/precip_jun_res_tgauss_sd_cb.tex}
\includegraphics[height=25mm,bb= 55  636    102  778,clip=]{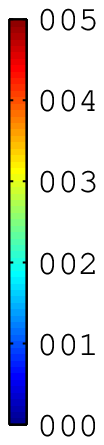}%
\end{minipage}
\end{center}
\vspace{-0.4cm}
\caption{Local estimates of the kriging residual standard deviations for the transformed Gaussian and GAL models using PRISM as a covariate.}
\label{fig:residuals}
\end{figure}

Figure \ref{fig:residuals} shows local estimates of the standard deviation of the residuals for the GAL and transformed Gaussian models. Clear spatial structures are seen in the figure for both models, indicating that neither model manage to use all spatial information contained in the observations. 

In the light of these results, it is interesting that the cross-validation results were not that different between the models, at least for the June data, and this indicates that the cross-validation procedure we used might not be appropriate for selecting the best model. However, the transformed Gaussian model, on it's ideal grid, would preform best overall, which indicates that the model variance needs to be non-stationary and proportional to the mean.

\section{Conclusions}\label{sec:conclusions}
In this work, we have extended the models of \cite{bolin11} to a larger class of non-gaussian models and have shown how to handle both missing data and measurement noise, which is crucial for practical implementations. 

The models and the estimation procedure can be extend and improved in several directions. For example, as previously mentioned, for models defined on regular grids the full generalized hyperbolic class could be used and thus give a very large class of non-Gaussian fields on lattices. Also, the estimation procedure was derived assuming that the field was observed under Gaussian measurement error, but it would require only small modification to extend it to Generalised hyperbolic measurement noise.

It is well-known that the convergence of the EM algorithm is slow, which often means that a large number of iterations are needed to achieve convergence of the parameter estimates, and the algorithm in this article is no exception. The author plans to study other stochastic estimation methods to improve the speed of the estimation. Changing to other estimation methods could also solve the problem that we are currently only able to estimate the parameters when $\alpha$ is an even integer. 

Unlike for Gaussian models, the models described here are not completely determined by the mean and covariance structures. This allows for interesting characteristics when applying other PDEs to the G-type processes.
For example, one can create a spatio-temporal model that is not time reversible by considering a spatio-temporal extension of the models discussed in this work.

For the precipitation data, the results indicate that the model should allow for kriging variances proportional to the kriging predictions, as the transformed Gaussian model does. This means that one needs to find ways to extend the models presented here to incorporate non-stationary variances.

\bibliographystyle{imsart-nameyear}
\bibliography{levybib}%

\end{document}